\documentclass[twocolumn,showpacs,preprintnumbers,amsmath,amssymb]{revtex4}

\usepackage{graphicx}
\usepackage{dcolumn}
\usepackage{bm}

\begin{document}


\title{Particle-hole symmetry and transport properties of the flux state
\\
in underdoped cuprates}
%
\author{Masaru Onoda$^1$}
\email{m.onoda@aist.go.jp}
\author{Naoto Nagaosa$^{1,2}$}
\email{nagaosa@appi.t.u-tokyo.ac.jp}
\affiliation{
$^1$Correlated Electron Research Center (CERC),
National Institute of Advanced Industrial Science and Technology (AIST),
Tsukuba Central 4, Tsukuba 305-8562, Japan\\
$^2$Department of Applied Physics, University of Tokyo,
Bunkyo-ku, Tokyo 113-8656, Japan
}
%
%
\date{\today}
%
\begin{abstract}
Transport properties, i.e., conductivities $\sigma_{\mu \nu}$, 
Hall constant $R_H$, and thermopower $S$ are studied for the flux state with 
the gauge flux $\phi$ per plaquett, which may model the underdoped cuprates, 
with the emphasis on the particle-hole and parity/chiral symmetries.
This model is reduced to the Dirac fermions in (2+1)D
with a mass gap introduced by the antiferromagnetic (AF) long range order 
and/or the stripe formation.
Without the mass gap, the Hall constant $R_H$ and
the thermopower $S$ obey the two-parameter scaling laws,
$R_H \cong \frac{a^2}{|e|x}
f_{R_H}\left(\frac{t\sqrt{x}}{k_BT}, \frac{\hbar}{\tau k_B T}\right)$ and
$S \cong \frac{k_B}{|e|}f_{S}\left(\frac{t\sqrt{x}}{k_BT}, 
\frac{\hbar}{\tau k_B T}\right)$,
with $a$ being the lattice constant, $x$ the hole concentration, 
$\tau$ the transport lifetime.
The $R_H$ and $S$ show the strong temperature dependence due to the 
recovery of the particle-hole symmetry at high temperature. 
The $x$-dependences of $\sigma_{xx} 
(\propto \sqrt{x})$ and $\sigma_{xy}$ (independent of $x$) 
are in a sharp contradiction with the experiments.
Therefore there is no signature of the particle-hole symmetry
or the massless Dirac fermions in the underdoped cuprates even above 
the Neel temperature $T_N$. 
With the mass gap introduced by the AF order, there occurs the 
parity anomaly for each of the Dirac fermions. However the 
contributions from different valleys and spins cancel with each other
to result in no spontaneous Hall effect even if the time-reversal
symmetry is broken with $\phi \ne \pi$.
The effects of the stripes are also studied.
The diagonal and vertical (horizontal) stripes have quite different influence 
on the transport properties.
The suppression of $R_H$ occurs at low temperature only when 
(i) both the AF order and the vertical (horizontal) stripe coexist, and 
(ii) the average over the in-plane direction is taken. 
Discussions on the recent experiments are given from the 
viewpoint of these theoretical results.
\end{abstract}
\pacs{74.25.Fy, 74.72.-h, 71.10.Fd, 72.15.Eb}
\maketitle

\section{Introduction}
Since the discovery of high-Tc cuprates, intensive studies have been done on 
the two-dimensional antiferromagnets. It is now established that the 
ground state of the 2D Heisenberg antiferromagnet with the nearest neighbor 
interaction on the square lattice shows the antiferromagnetic (AF) 
long range ordering at zero temperature\cite{Neel}, 
and the low energy spin excitation can be described in terms of 
the spin wave theory. However this does not mean that the electronic state in 
the antiferromagnets is fully understood.  Compared with the triplet channel
of the two-particle correlation functions, the single particle properties such 
as the angle resolved photo-emission spectra (ARPES)\cite{ARPES}, 
and the singlet channel correlation functions such as the charge transport 
properties\cite{LSCO-RH,LSCO-rho,YBCO-rho,PH-symmetry,ando1,ando2} 
still remain controversial.
In fact, there are two different pictures for it.  
One is the conventional spin density wave (SDW) picture\cite{spin-bag} 
with the wavenumber ${\vec Q} = (\pi,\pi)$ where both 
the weak and strong coupling regions can be smoothly connected. 
The other picture is the $\pi$-flux state\cite{SFP} 
originated from the resonating valence bond (RVB) idea\cite{RVB}.
At half filling, the $\pi$-flux state is equivalent to 
the d-wave singlet paring state\cite{dWP,sRVB} due to 
the particle-hole SU(2) symmetry\cite{affleck}.
The Gutzwiller projected wavefunction with the d-wave paring and AF orders 
gives a better energy compared with that only with the AF order\cite{AF-dSC}. 
Also the higher energy continuum of the neutron scattering spectra and 
Raman scattering spectra has been analyzed in terms of this $\pi$-flux 
state\cite{AF-dSC}.

The dispersion of a single hole put into this antiferromagnet is 
an important issue studied intensively in terms of 
self-consistent Born approximation\cite{scba}, 
exact diagonalization\cite{ed}, spinon-holon bound state picture\cite{s-h}, 
and variational method\cite{vari}.
With the $t$-$J$ model, all the analyses give the maximum of the
hole dispersion at ${\vec k} = (\pm \pi/2, \pm \pi/2)$.
This dispersion can be understood in terms of the 
$\pi$-flux  picture\cite{s-h,vari}, which 
introduces the nodal Fermi points of the spinons at 
${\vec k} = (\pm \pi/2, \pm \pi/2)$ with the dispersion similar to the 
d-wave superconductors. 
This fits the ARPES experiments in the undoped cuprates\cite{ARPES}.

At finite doping, the slave-boson mean-field theory of the $t$-$J$ model 
predicts the state with both the singlet RVB and AF orders 
for small $x$\cite{AF-sRVB}.
The SU(2) symmetry has been employed to represent the constraint 
and the underdoped pseudogap region is characterized 
as the staggered flux state with spin-charge separation\cite{SU2}, 
which can be regarded as the fluctuating state between 
the d-wave pairing state and the current order state. 
On the other hand, the staggered flux state with the electron coordinates 
with the real current ordering and periodicity doubling has been proposed 
for the underdoped cuprates\cite{hidden}.

Recently there appeared several experiments on the Hall coefficient $R_H$  
and the thermopower $S$ in the heavily underdoped cuprates, which raised the 
issue of particle-hole symmetry\cite{PH-symmetry,ando1,ando2}. 
The $R_H$ as well as the $S$ show the strong suppression below the Neel 
temperature $T_N$ in some YBCO samples\cite{PH-symmetry} while it is
not the case for other samples\cite{ando1}.  This strongly suggests that the 
transport properties below $T_N$ is sensitive to the oxygen chain
ordering and/or the self-organization of electrons such as the 
stripes, which depends on the annealing procedure in the sample preparation.
Another interesting clue is that in LSCO the suppression of the Hall effect 
is observed only in the vertical (horizontal) stripe state while is not in the
diagonal stripe state for $x<0.05$\cite{noda,uchida}. Above $T_N$,
$R_H$ is a decreasing function of temperature, 
which remains one of the most 
puzzling features in the normal state 
properties\cite{LSCO-RH,YBCO-rho,conseq,PH-symmetry}.

In the SDW picture, the large metallic Fermi surface enclosing the 
area of $1-x$ ($x$: hole concentration) is recovered above $T_N$. 
Therefore we expect the drastic change of the Hall constant
from $R_H = -a^2/(1-x)$ to $R_H = a^2/x$
when the temperature $T$ is lowered across $T_N$. 
(Here $a$ is the lattice constant.)
Correspondingly the resistivity should be
affected by the onset of the AF order.
This is in sharp contradiction with the experiments.
Even above $T_N$, the system behaves as the doped Mott insulator 
with the small number of holes\cite{LSCO-RH,YBCO-rho,PH-symmetry}, 
and even a slight change of the resistivity is not
observed at $T_N$\cite{ando1}.

Considering all these clues, 
it is worthwhile to study the transport properties of the slightly doped 
$\pi$-flux state both above and below $T_N$, which we undertake in this paper. 
We will neglect the interactions between electrons and the disorder potentials.
The former is justified because the quasi-particle number is small 
at low temperature due to the reduced density of states and 
the el-el interaction is irrelevant. 
The latter becomes important at low temperature for small $x$ 
where the resistivity shows an upturn, but is irrelevant 
for the temperature and $x$-range of our interest.  
We also study the effect of the stripe formation on the Hall constant
and the thermopower, and show that oxygen chain ordering is crucial for these
quantities.  The deviation of the flux from $\pi$, which breaks 
the time-reversal symmetry and produces current ordering, is also studied.
This issue is closely related to the parity anomaly\cite{Anomaly} 
in (2+1)D because the two species of the Dirac fermions 
acquires the mass gap due to the AF ordering.
Therefore the undoped and underdoped cuprates offers an interesting 
laboratory to study the transport properties of the Dirac fermions
with nontrivial topological nature influenced by the AF order and/or the 
stripe formation. 
The Dirac fermion has been studied also in the context of 
the nodal quasi-particle in the d-wave supercouductors\cite{nodal}. 
In this case the Fermi
energy is always at $E=0$ and the particles-hole symmetry remains. 
Under the external magnetic field $H$, the formation of the vortex
lattice is crucial, which introduces the magnetic length scale 
$\ell_H \sim H^{-1/2}$. In the present case, on the other hand, 
there occurs no Meissner effect, and one can see the response of the 
Dirac fermions in the uniform state, while the particle-hole symmetry is 
broken due to the shift of the Fermi energy from $E=0$. This introduces 
the length  scale $\ell_x \sim x^{-1/2}$ , i.e.,
the inter-hole distance, and we can have the new kind of scaling 
when neither AF order nor the stripe formation occur.

The plan of this paper follows. In sec.~\ref{sec:hamiltonian}, 
our model Hamiltonian is 
introduced and its spectrum under the external magnetic field is reviewed. 
In sec.~\ref{sec:responses}, 
the electromagnetic and thermal responses are 
studied for the Dirac fermions without stripes.
The effects of the stripes are studied in sec.~\ref{sec:stripe}, 
and sec.~\ref{sec:discussion} is 
devoted to discussions in comparison with experiments.

\section{\label{sec:hamiltonian}Hamiltonian}
We start with the most generic Hamiltonian in the staggered flux state
with anti-ferromagnetic and stripe (quasi-)order
which are treated as on-site potentials in the  mean field theory.
Hereafter, we take the units in which $\hbar=c=1$.
\begin{eqnarray}
H 
&=& -\sum_{\langle \vec{r},\vec{r}'\rangle, \, \sigma} 
t_{\vec{r},\vec{r}'}
c_{\vec{r} \sigma}^{\dagger}c_{\vec{r}' \sigma}
\nonumber\\
&-&\sum_{\vec{r}, \,\sigma }\left[\mu  
+ u_\sigma  \cos(\vec{Q}\cdot \vec{r})+v\cos(\vec{Q}_{v}\cdot \vec{r})\right]
c_{\vec{r} \sigma}^{\dagger}c_{\vec{r} \sigma}.
\label{eq:Hamiltonian}
\end{eqnarray}
The transfer integral 
$t_{\vec{r}\pm\hat{x},\vec{r}} = t_{\vec{r}\pm\hat{y},\vec{r}}^* 
= t e^{i\frac{\phi}{4}} $
(with $r_x+r_y= \mbox{even}$) 
represents the staggered flux $\phi$ for each plaquette. 
In the RVB picture, $\phi$ is generated by 
the superexchange interaction $J$ and equals $\pi$ in the undoped case.
Therefore, in our representation,
the transfer integral $t$ is estimated as $t \cong 0.8J$,
and should not be confused with $t$ in the $t$-$J$ model\cite{s-h}.
In the second term,  $u_\sigma$ is the mean field potential for the 
$\sigma$ spin electrons with the
wavenumber $\vec{Q} = (\pi,\pi)$, which originates from the AF order $m_{AF}$ 
and/or the diagonal stripe formation $m_{\rm diag}$ as 
$u_\sigma = \sigma m_{AF} + m_{\rm diag}$, while
$v$ is that from the vertical stripe formation with $\vec{Q}_v = (\pi,0)$.
This is the result of the higher order process in the stripe potential with 
the wavenumber $\vec{Q}_{\rm stripe} = (\pi/M, \pi/M)$ (diagonal case) or 
$\vec{Q}_{\rm stripe} = ( \pi/M, 0)$ (vertical case).
The chemical potential $\mu$ is zero at the half-filling.
The eigenvalues for each $\vec{k}$ are 
\begin{widetext}
\begin{equation}
\epsilon_{\sigma}(\vec{k}) =
\pm\Biggl[
4t^2 (\cos^2{k_x}+\cos^2{k_y})+u_{\sigma}^2+v^2
\pm 2\sqrt{
\left(4t^2\cos\frac{\phi}{2}\cos{k_x}\cos{k_y}\right)^2
+v^2\left(4t^2\cos^2{k_y}+u_{\sigma}^2\right)}
\Biggr]^{\frac12}.
\label{eq:eigenvalue}
\end{equation}
\end{widetext}

First, we consider the case without the vertical stripe. 
In this case, the low energy electronic states are described in terms of the 
two Dirac fermions at ${\vec k}_1 = (\pi/2,\pi/2)$ and 
$\vec{k}_2 =(\pi/2, - \pi/2)$, which we call 1 and 2 sector, respectively.
Measuring the momenta from these ${\vec k}_{1,2}$-points, and in the 
continuum approximation the Hamiltonian is given as
\begin{equation}
H_{\rm eff} = \sum_{i, \sigma}
\int d^2r\;
\mbox{\boldmath$\psi$}_{i \sigma}^{\dagger}(\vec{r})
\hat{\cal H}_{i \sigma}
\mbox{\boldmath$\psi$}_{i \sigma}(\vec{r}),
\label{eq:H_eff}
\end{equation}
where $\mbox{\boldmath$\psi$}_{i \sigma}^{(\dagger)}(\vec{r})$ is 
the annihilation (creation) operator of 
$i$-sector ($i=1,2$) with $\sigma$-spin,  
\begin{equation}
\hat{\cal H}_{1 \sigma}
=\left[
\begin{matrix}
u_{\sigma} - \mu & 2ta(\hat{p}_x+e^{-i\frac{\phi}{2}}\hat{p}_y)\\
2ta(\hat{p}_x+e^{i\frac{\phi}{2}}\hat{p}_y)& -u_{\sigma} - \mu
\end{matrix}
\right],
\label{eq:1st-Dirac}
\end{equation}
and 
$\hat{\cal H}_{2  \sigma} = \hat{\cal H}_{1 \sigma}
\Bigr|_{{u_{\sigma} \to -u_{\sigma}}\atop{\phi\to2\pi-\phi}}$.
Here $\hat{p}_{\mu}=-i\partial_{\mu}$ and $a$ is the lattice constant.
As is evident from the above Hamiltonian, each Dirac fermion has
the mass term with positive or negative sign, 
and shows the parity anomaly\cite{Anomaly}.
When the external electromagnetic (EM) field $A_\mu$ is coupled 
to each Dirac fermion, the Chern-Simons term
$\varepsilon^{\mu \nu \lambda} A_\mu \partial_\nu A_\lambda $
is generated. Therefore the Dirac fermion with positive (negative)
$u_{\sigma}$ has the right (left) chirality.
Because there are two right (R) and left (L) Dirac fermions,
there occurs the cancellation of the Chern-Simons terms, and no
spontaneous Hall effect results.
This remains true even when the flux $\phi$ is different from $\pi$
and the time-reversal symmetry is broken to produce the current order,
as long as diagonal stripes are absent, i.e. $m_{\rm diag}=0$.
It is because the current pattern is staggered, and does not
affect the uniform response in an essential way.

Now we briefly review on the Dirac fermions 
in the presence of the external uniform magnetic field $B$,
which can be analytically solved\cite{3D-Dirac-B,2D-Dirac-B}.
In the effective theory represented by eq.~(\ref{eq:H_eff}),
the first-quantized Hamiltonian eq.~(\ref{eq:1st-Dirac})
is rewritten by the replacement,
$\hat{p}_{\mu} \to \hat{\pi}_{\mu} 
= \hat{p}_{\mu} - e A_{\mu}(\vec{r})$,
where $[\nabla\times \vec{A}(\vec{r})]_{z} = B$.
Then we define the following bosonic operators as
\begin{eqnarray}
&&\hat{a}_{\phi}=\frac1{\sqrt{2|eB_{\phi}|}}
\left[\hat{\pi}_x+e^{i\,\mbox{sgn}(eB_{\phi})\frac{\phi}{2}}
\hat{\pi}_y \right],
\\
&&\hat{a}_{\phi}^{\dagger}=\frac1{\sqrt{2|eB_{\phi}|}}
\left[\hat{\pi}_x+e^{-i\,\mbox{sgn}(eB_{\phi})\frac{\phi}{2}}
\hat{\pi}_y \right].
\end{eqnarray}
These operators satisfy the commutation relation 
$[\hat{a}_{\phi}, \hat{a}_{\phi}^{\dagger}]=1$.
Using these operators and the Pauli matrices $\tau_i$, ($i=1,2,3$), 
eq.~(\ref{eq:1st-Dirac}) is rewritten as
\begin{eqnarray}
\hat{\cal H}_{1 \sigma}
&=&\frac{K_{eB_{\phi}}}{2}
\left[
(\hat{a}_{\phi}+\hat{a}_{\phi}^{\dagger})\,\tau_1
-i\,\mbox{sgn}(eB_{\phi})(\hat{a}_{\phi}-\hat{a}_{\phi}^{\dagger})\,\tau_2
\right]
\nonumber\\
&&+u_{\sigma}\,\tau_3 -\mu I.
\label{eq:1st-Dirac-B}
\end{eqnarray}
The eigenvector ($n\ge 1$) is
\begin{widetext}
\begin{equation}
|n\,\sigma\,\pm\rangle
=\frac1{\sqrt{2\epsilon_n}}
\left[
\sqrt{\epsilon_n\pm \mbox{sgn}(eB_{\phi})u_{\sigma}}
\,|n\rangle\otimes|\tilde{\uparrow}\rangle
\pm \sqrt{\epsilon_n\mp \mbox{sgn}(eB_{\phi})u_{\sigma}}
\,|n-1\rangle\otimes|\tilde{\downarrow}\rangle
\right]\otimes|\sigma\rangle,
\label{eq:eigenvector}
\end{equation}
\end{widetext}
with the eigenvalue $\pm\epsilon_{n\sigma}$ being given by
\begin{eqnarray}
&&\epsilon_{n\sigma} = \sqrt{K_{eB_\phi}^2 n +u_{\sigma}^2},
\\
&&K_{eB_\phi} = \sqrt{8t^2a^2|eB_\phi|}, \quad
B_\phi = B \sin\frac{\phi}{2}.
\end{eqnarray}
Here $|n\rangle$ is the eigenvector of 
$\hat{a}_{\phi}^{\dagger}\hat{a}_{\phi}$ with the eigenvalue $n$,
$|\tilde{\uparrow}(\tilde{\downarrow})\rangle$ is the eigenvector of 
$\mbox{sgn}(eB_{\phi})\tau_3$ with the eigenvalue $+1(-1)$,
and the quantum index for the intra-level orbitals
is omitted.
$|\tilde{\uparrow}(\tilde{\downarrow})\rangle$
should not be confused with a real-spin eigenvector $|\sigma\rangle$,
but it comes from the two-component nature of a Dirac fermion.
Especially, the zero mode is given by
\begin{equation}
|0\,\sigma\,\pm\rangle
=\frac12
\left[1\pm\mbox{sgn}(u_{\sigma}eB_{\phi})\right]
|0\rangle\otimes|\tilde{\uparrow}\rangle
\otimes|\sigma\rangle.
\label{eq:zero-mode}
\end{equation}
Therefore, there exists only one zero-mode $|0\,\sigma\,+\rangle$ 
or $|0\,\sigma\,-\rangle$ for each sector and its energy is 
$\epsilon_ {0\sigma} = |u_{\sigma}| $ or $-|u_{\sigma}|$ 
depending on the direction of 
the external magnetic field $B$ and the chirality of Dirac fermions.
We can also define effective current-operators as follows,
$\hat{j}_x=2ta\,\tau_1$ and 
$\hat{j}_y=2ta(\cos{\frac{\phi}2}\,\tau_1+\sin{\frac{\phi}2}\,\tau_2)$

\section{\label{sec:responses}Electromagnetic and thermal responses}
The electromagnetic and thermal linear response functions
are obtained by the Kubo formula.
We put the above solutions into the Kubo formula 
for $\sigma_{\mu \nu}(\omega)$ and approximate the effect of the 
relaxation by replacing $\omega$ by $\omega + i/\tau$
with the lifetime $\tau$. This approximation reproduces the 
Drude formula for $\sigma_{\mu \nu}(\omega)$ in the simplest case. 
The contribution from $1$-sector with $\sigma$-spin is presented 
in the appendix~\ref{sec:conductivity}.
The total conductivity is given by summing up the contribution
from all sectors.
This procedure makes anomalous terms,
which come from the inter-band effect,
cancel out with each other,
and lead to the consistent result in the AF ordered state,
i.e. no parity symmetry breaking in the limit $B\to 0$.

The eqs.~(\ref{eq:sigma_xx}) and (\ref{eq:sigma_xy}) are valid 
for general cases including both 
the quantum limit ($\tau \to \infty$ with finite $B$) and 
the semi-classical limit 
($K_{eB_\phi}\tau \ll 1 \ll |\mu|\tau$).
In the former case, eqs.~(\ref{eq:sigma_xx}) and (\ref{eq:sigma_xy})
represent the integer quantum Hall effect, 
while in the latter case they correspond to the usual Hall effect.
In the latter case, 
we obtain the same formula with that given by the Boltzmann equation.
For example, with only the AF order and at low temperature, 
eqs.~(\ref{eq:sigma_xx}) and (\ref{eq:sigma_xy}) lead to 
\begin{eqnarray}
\sigma_{xx} 
&\cong&
\frac{e^2}{2\pi}\cdot 
\frac{8\pi x \,t^2\tau}
{\sqrt{4\pi x \,t^2\left|\sin{\frac{\phi}{2}}\right|+m_{AF}^2}},
\label{eq:sigma_xx_low}
\\
\sigma_{xy}
&\cong&
\frac{e^2}{2\pi}\cdot
\frac{(8\pi x \,t^2 \tau)^2}
{4\pi  x \,t^2\left|\sin{\frac{\phi}{2}}\right|+m_{AF}^2}
\cdot\frac{a^2 |e|B}{2\pi x}\sin^2{\frac{\phi}{2}},
\label{eq:sigma_xy_low}
\end{eqnarray}
and the Hall coefficient $R_H$ is given by
\begin{equation}
R_H \cong
{ \frac{\sigma_{xy}}{B \sigma_{xx}^2} }
 \cong \frac{a^2}{|e| x}\sin^2{\frac{\phi}{2}}.
\label{eq:RH-SFP}
\end{equation}
independent of the presence/absence of the AF order.
This is reduced to the conventional result 
$R_H = a^2/(|e| x)$ for the hole system with concentration 
$x$ when we put $\phi=\pi$. 
However the $x$-dependences of $\sigma_{xx}$
and $\sigma_{xy}$ are peculiar in the absence of the 
AF order, namely $\sigma_{xx} \propto \sqrt{x}$
and $\sigma_{xy}$ being independent of $x$
when we neglect the $x$-dependence of $\tau$.
These are understood as follows.
The $\sigma_{xx}$ is proportional to the density of states
at the Fermi energy, which is proportional to $\sqrt{x}$
for the massless Dirac fermion.
As for $\sigma_{xy}$, on the other hand,
the geometric interpretation is useful\cite{geometric}.
The Hall conductivity is determined by the  scattering path length,
$\vec{l}(\vec{k})=\tau \vec{\nabla}\epsilon(\vec{k})$, 
which is independent of the magnitude $|\vec k|$ for linear
dispersion and hence $\sigma_{xy}$ is independent of $x$.
In eq.~(\ref{eq:sigma_xy}), this behavior is the consequence of the dominant 
contribution from the zero mode of the Dirac fermions to $\sigma_{xy}$.
On the other hand, when $\sqrt{4\pi x}\,t\ll |m_{AF}|$, the usual dispersion 
$\epsilon(\vec{k}) \propto \vec{k}^2$ is relevant and 
the usual $x$-dependences, 
$\sigma_{xx} \propto x$ and $\sigma_{xy} \propto x$ result.
The thermopower $S$ at sufficiently low temperature
is also obtained in a similar way as
\begin{equation}
S \cong
\frac{k_B}{|e|}\cdot \frac{\pi^2}{3}k_B T\cdot
\frac{4\pi  x \,t^2 \left|\sin{\frac{\phi}{2}}\right|+2m_{AF}^2}
{4\pi  x \,t^2 \left|\sin{\frac{\phi}{2}}\right|
\left[4\pi x \,t^2\left|\sin{\frac{\phi}{2}}\right|+m_{AF}^2
\right]^{\frac12}}.
\label{eq:S-SFP}
\end{equation}
The $x$-dependence of $S$ is also peculiar in the absence of the AF order,
i.e., $S\propto 1/\sqrt{x}$.

\begin{figure*}
  \begin{tabular}{cc}
    \includegraphics[scale=0.4]{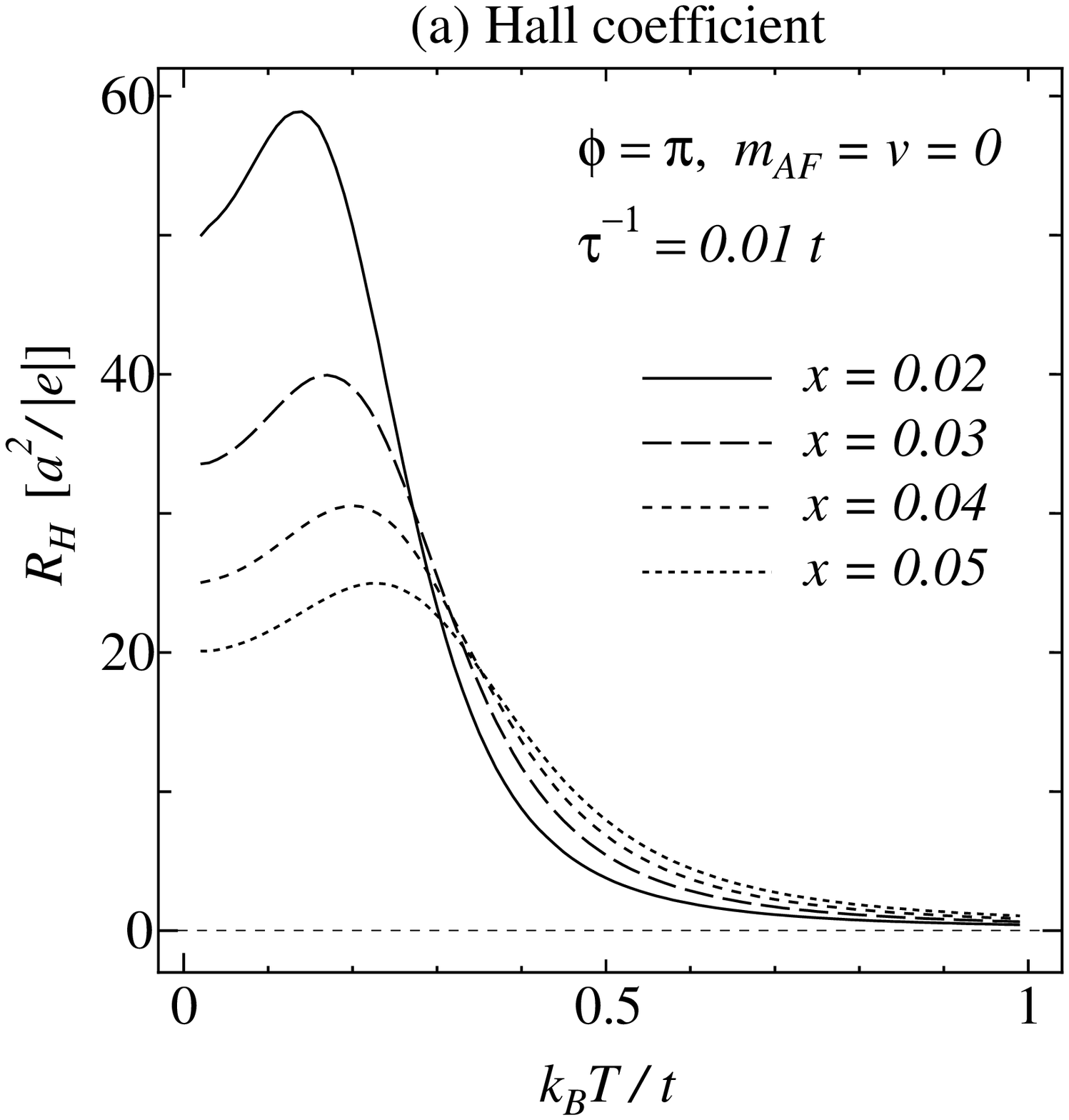} &
    \includegraphics[scale=0.4]{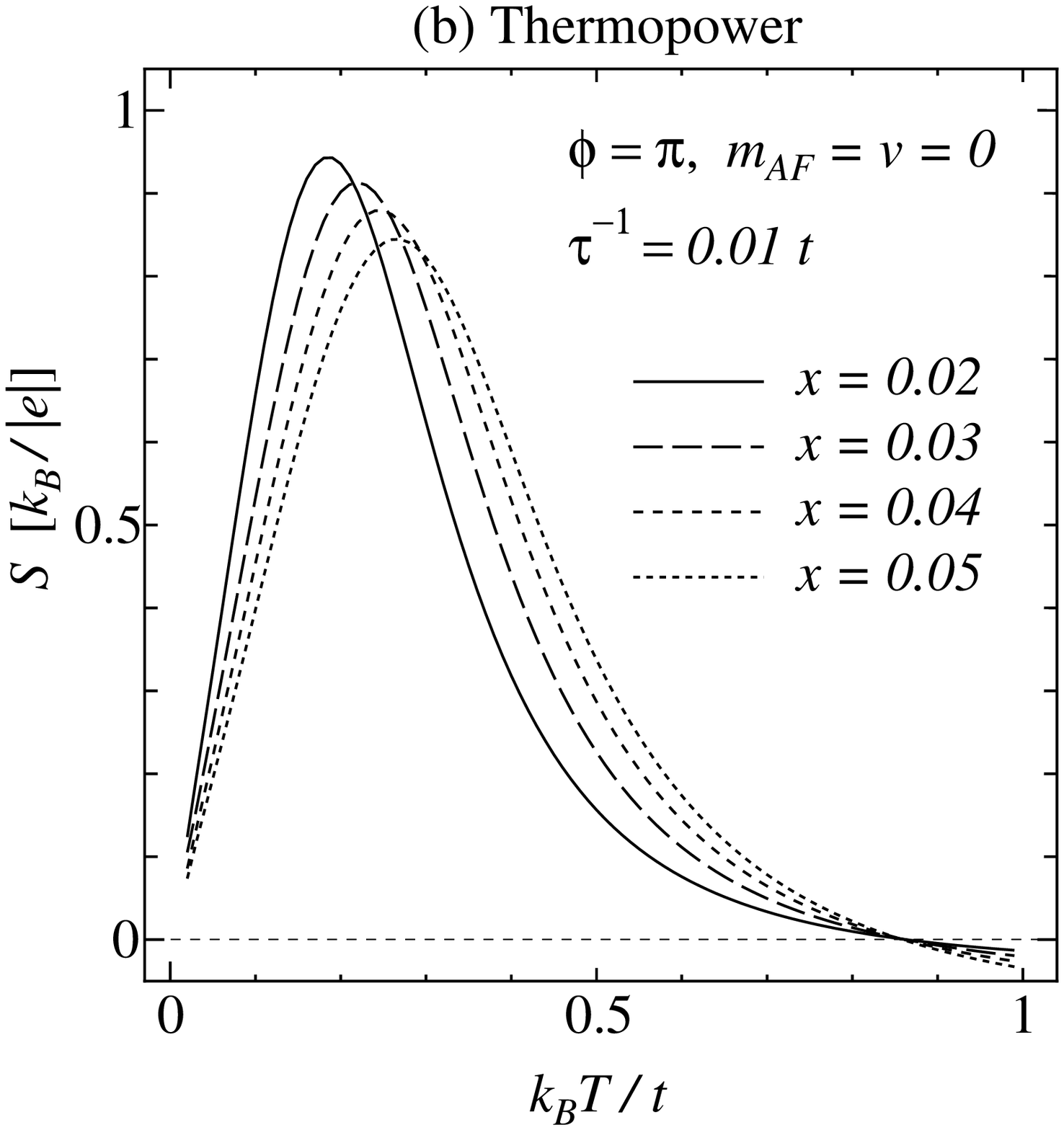} 
  \end{tabular}
  \caption{
Temperature dependences of
(a) the Hall coefficient $R_H$ in the unit $a^2/|e|$, and 
(b) the thermopower $S$ in the unit $k_B/|e|$
for $m_{AF}=v=0$.
}
\label{fig:RH-S.tmp}
\end{figure*}
Next we consider the temperature dependences of $R_H$ and $S$,
which originates from the Fermi distribution function 
$f_F(T, \epsilon, \mu(T))$.  
Here the energy is averaged over $\sim k_{B}T$, 
and both the particle and hole branches contribute 
with the opposite signs to $R_H$ and $S$
when $k_{B}T >|\mu(T)|$.
Furthermore, $|\mu(T)|$ is a decreasing function of temperature
as presented in the appendix~\ref{sec:mu-tmp}.
For example, when $m_{AF}=0$, it behaves at $k_B T \ll {\cal O}(t\sqrt{x})$ 
\begin{equation}
|\mu(T)| \cong 
\sqrt{4\pi x\, t^2 \left|\sin{\frac{\phi}2}\right|
-\frac{\pi^2}{3}(k_B T)^2},
\end{equation}
and at $k_B T \gg {\cal O}(t\sqrt{x})$
\begin{equation}
|\mu(T)| \cong 
\frac{\pi x\, t^2}{\ln 2}\left|\sin{\frac{\phi}2}\right|
\frac1{k_B T}.
\end{equation}
Therefore, as long as the flux order is present,
the particle-hole symmetry is approximately restored 
at high temperature even in the doped system,
and both $R_H$ and $S$ are reduced
as shown in Figs.~\ref{fig:RH-S.tmp}(a) and \ref{fig:RH-S.tmp}(b),
which are the results of the original lattice-model.
The peak in $R_H$ occurs because the suppression of $\sigma_{xx}$ as 
increasing $T$ dominates at low temperature.
After all, the low carrier density and the narrow gap
result in these features, 
i.e., having a peak at moderate temperature
and decreasing at high temperature.  
It is worthwhile to note that the temperature dependence of $S$
resembles those of rare-earth compounds, such as YbCuAl,
which have low carrier densities and narrow gaps\cite{mahan}.

\begin{figure*}
  \begin{tabular}{cc}
    \includegraphics[scale=0.4]{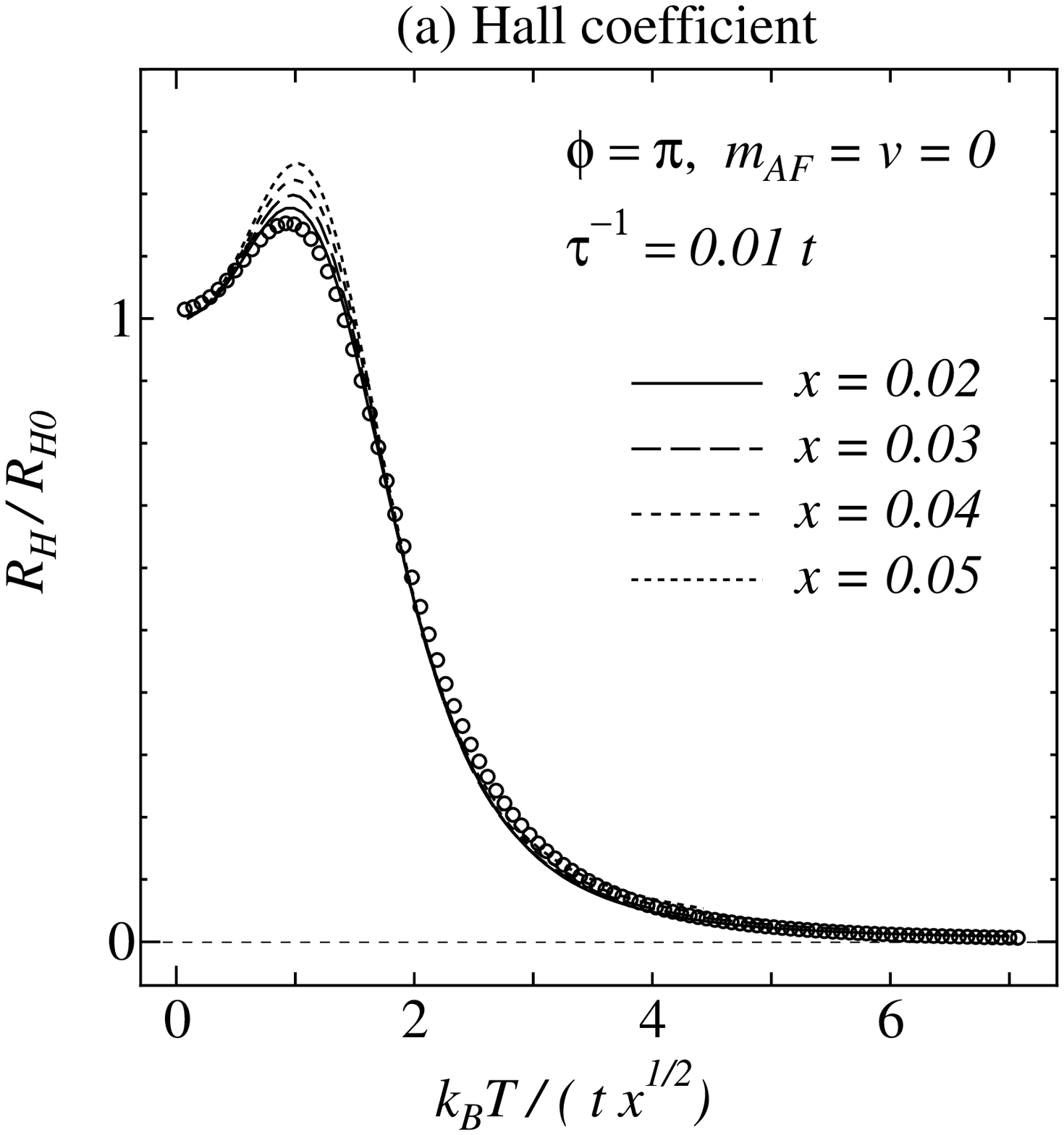} &
    \includegraphics[scale=0.4]{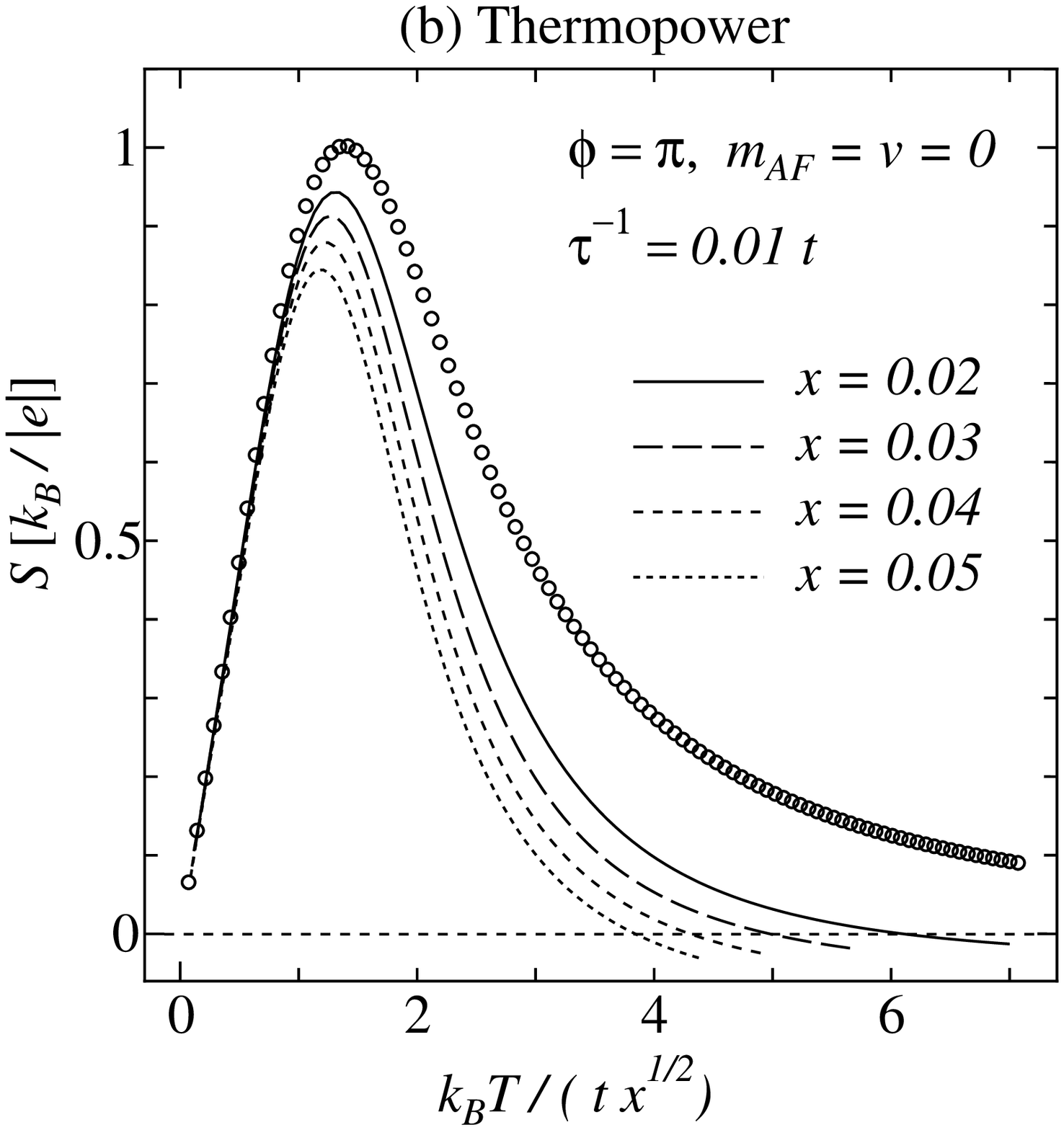} 
  \end{tabular}
  \caption{
Scaling of temperature dependences of
(a) the Hall coefficient $R_H$ divided by $R_{H0} = a^2/(|e|x)$, and 
(b) the thermopower $S$ in the unit $k_B/|e|$
for $m_{AF}=v=0$.
The open circles are scaling functions
calculated in the continuum model.
}
\label{fig:RH-S.scl}
\end{figure*}
In the case of $m_{AF} = 0$ and the limit $B\to 0$,
the following scaling laws are expected by the continuum model
\begin{eqnarray}
R_H &=& \frac{a^2}{|e|x}
f_{R_H}\left(\frac{t\sqrt{x}}{k_B T}, \frac1{\tau k_B T}\right),
\\
S &=& \frac{k_B}{|e|}
f_{S}\left(\frac{t\sqrt{x}}{k_B T}, \frac1{\tau k_B T}\right),
\end{eqnarray}
where $f_{R_H}$ and $f_{S}$ are dimensionless functions.
(The details are given in the appendix~\ref{sec:scaling}.)
The dependence on $t\sqrt{x}/(k_B T)$ is 
because of $\mu/(k_B T)$ being a dimensionless function
of $t\sqrt{x}/(k_B T)$, 
and the dependence on $\tau k_B T$ comes from 
the transition between the highest valence band
and the lowest conduction band.
The latter is neglected in the Boltzmann transport theory.
In a physical sense, the contribution from transitions between two bands 
would be very small when the condition $1 \ll |\mu| \tau$ is satisfied,
i.e. in the semi-classical limit.
In this case, we get the one-parameter scaling lows,
\begin{equation}
R_H \cong \frac{a^2}{|e|x}f_{R_H}\left(\frac{t\sqrt{x}}{k_B T}\right),
\quad
S \cong \frac{k_B}{|e|}f_{S}\left(\frac{t\sqrt{x}}{k_B T}\right).
\end{equation}
Therefore, both $x R_H$ and $S$ are expected to scale as 
functions of $t\sqrt{x}/(k_B T)$.
Figs.~\ref{fig:RH-S.scl}(a) and \ref{fig:RH-S.scl}(b) show
the scaling behaviors of $R_H$ and $S$ 
given in Figs.~\ref{fig:RH-S.tmp}(a) and \ref{fig:RH-S.tmp}(b).
We can see that the single-parameter scaling law of $R_H$
works fairly well in broad temperature range.
On the other hand, for $S$, it works only at low temperature.
This is because $S$ contains the energy integral 
which has one more energy-dimension than the integral for $R_H$,
and $S$ is more sensitive to the higher energy region where 
the lattice structure is relevant.

\begin{figure*}
  \begin{tabular}{cc}
    \includegraphics[scale=0.4]{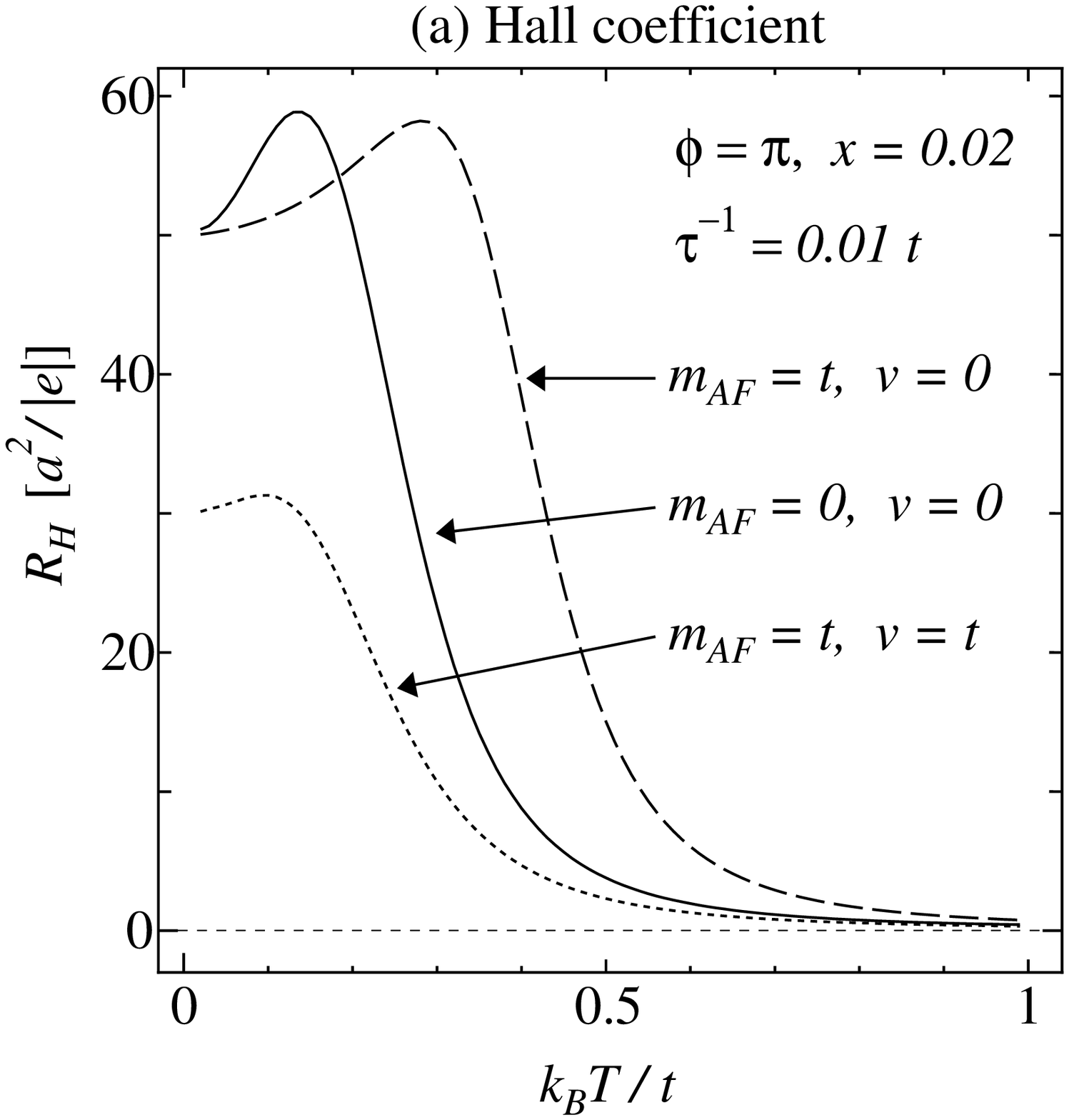} &
    \includegraphics[scale=0.4]{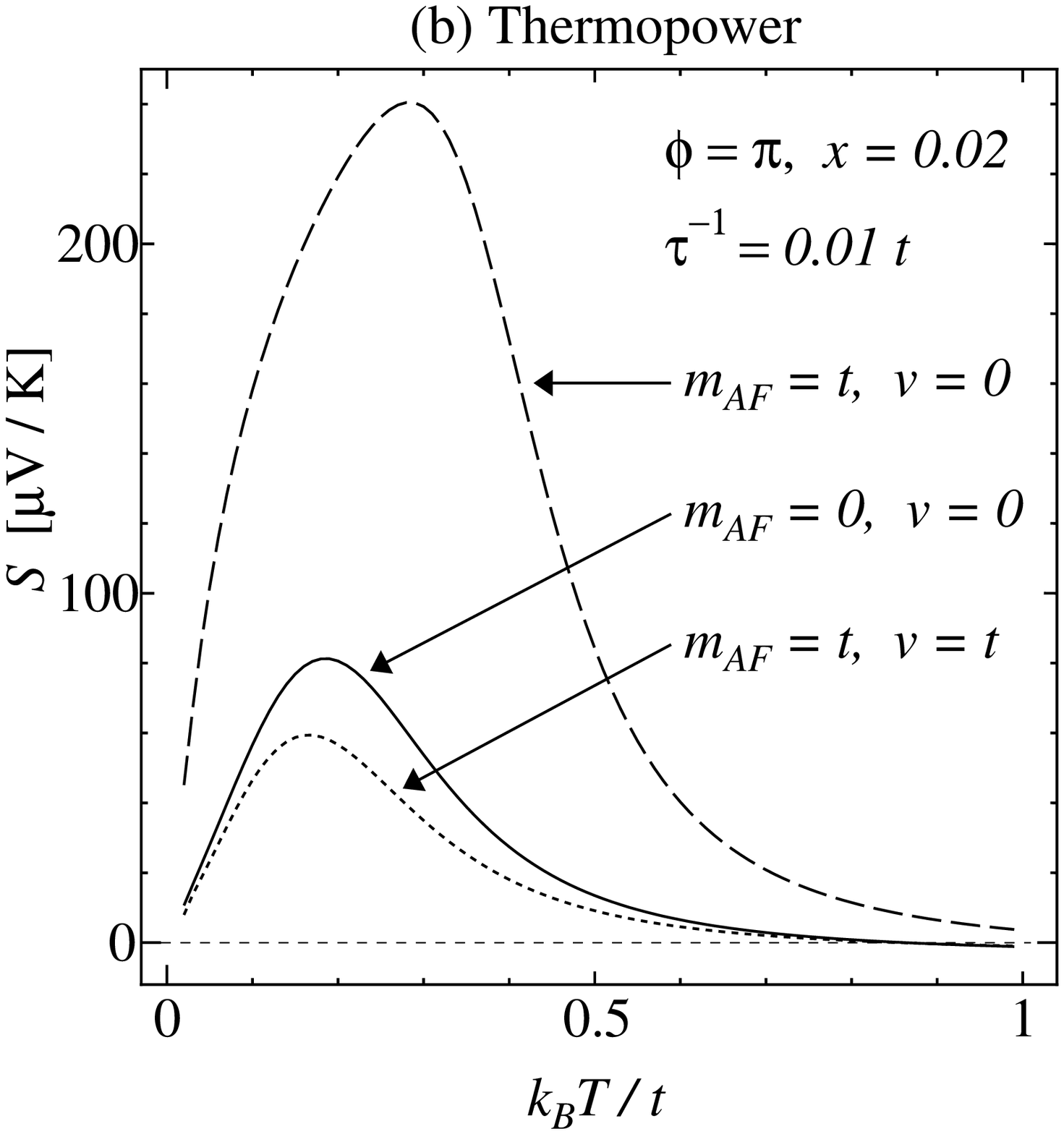} 
  \end{tabular}
  \caption{
Temperature dependences of
(a) the Hall coefficient $R_H$ in the unit $a^2/|e|$, and 
(b) the thermopower $S$ in the unit $\mu {\rm V/K}$
for $(m_{AF}, v) = (0, 0)$, $(t, 0)$ and $(t, t)$.
}
\label{fig:RH-S.tmp.u-v}
\end{figure*}

In Figs.~\ref{fig:RH-S.tmp.u-v}(a) and \ref{fig:RH-S.tmp.u-v}(b),
the temperature dependences of $R_H$ and $S$ are shown 
in the presence of the AF order and the vertical (horizontal) stripe.
These results are obtained for the lattice model
in terms of the Boltzmann transport theory 
where the inter-band effect is neglected.
We can observe that the AF order (without the stripe formation) 
results in the enhancement of $R_H$ and $S$.
One reason of this enhancement is that
the AF order suppresses the recovery of 
the particle-hole symmetry at high temperature.
(The details are given in the appendix~\ref{sec:mu-tmp}.)
Another reason is that 
the heavier mass $m_{AF}$ makes the conductivity smaller
as seen in eqs.~(\ref{eq:sigma_xx_low}) and (\ref{eq:sigma_xy_low}).
The former reason is crucial both for $R_H$ and $S$.
On the other hand, the latter would be crucial only for $S$
as expected from the comparison of eqs.~(\ref{eq:RH-SFP}) and (\ref{eq:S-SFP}).
This is because the explicit dependence of $\sigma_{xx}$ and $\sigma_{xy}$
on $m_{AF}$ nearly cancel each other in $R_H$.
Therefore, the enhancement of $S$ is more drastic than that of $R_H$. 
As for the effect of the stripe,
we will consider it closely in the next section.

\section{\label{sec:stripe}Effect of stripe formation}
The quasi-one dimensional spin/charge ordering 
occurs in some cuprates\cite{LNSCO-stripe-exp}, and consequently 
affects the transport properties\cite{ando1,ando2,noda}.
Therefore it is worthwhile to study the effect of 
the stripe formation on the flux state.
Some works have been done assuming the one-dimensionality, 
which corresponds the limit of strong stripe potential\cite{maekawa}. 
In this section, we give an alternative and complemental study starting 
from the 2D flux state.
It is easy to see that the effects of the stripes are essentially
different between the diagonal and vertical (horizontal) ones, because
the vertical (horizontal) stripe introduces the off-diagonal matrix elements
between the two Dirac fermions around ${\vec k}_1$ and 
${\vec k}_2$, while the diagonal one does not.
Therefore the effect of the diagonal stripe is similar to that of
the AF order and we do not expect any drastic change of 
$R_H$ in the case of the diagonal stripe,
although it makes the mass gaps of different spins unbalanced
and leads to the modification of $S$.
It is noted that the change of the mass gap modifies $S$
as in eq.~(\ref{eq:S-SFP}).
(The unbalance of the mass gaps of different spins
also leads to the ferrimagnetism.)
After all, the drastic modifications of both $R_H$ and $S$ are 
expected only when the vertical (horizontal) stripe formation occurs. 
We will focus on this case below. 

In the presence of the stripe, there appears the anisotropy in the plane.
A naive expectation is that the 1D nature of the transport 
along the stripe reduces $R_H$. 
However, this is not the case, 
because the stripe does suppress not only $\sigma_{xy}$ 
but also one of $\sigma_{xx}$ and $\sigma_{yy}$ 
with $R_H \cong \sigma_{xy}/( B \sigma_{xx} \sigma_{yy})$ being unchanged.
This is the case also in our explicit calculation showing that 
$R_H = a^2/(|e| x)$ is a robust feature at low temperature. 
At this point, we must consider the configuration of the stripe in cuprates.
When there are stripes, their direction would be different 
in each CuO$_2$ layer as in LSCO, or there are domains of stripes with 
different directions in a layer as in twined YBCO. In the former system, 
we simply assume that the vertical and horizontal types occurs alternately.
Then it is reasonable to average the contributions from different layers,
because they can be regarded as a {\it parallel} circuit,
i.e. we should take an average of conductivity tensor not 
of resistivity tensor.
Also in the latter system, the more conductive regions percolate and 
hence dominate the conductivity of the system. Therefore it is 
reasonable to take average of $\sigma_{xx}$ and $\sigma_{yy}$.
Then the observed $R_H$ is suppressed because only $\sigma_{xy}$ 
is reduced considerably while  $\sigma = (\sigma_{xx}+\sigma_{yy})/2$ is not,
and $R_H \cong \sigma_{xy}/(B \sigma^2)$ is suppressed.

Before discussing general cases,
we consider some special cases where $R_H$ and $S$
can be analytically evaluated at low temperature
by the Boltzmann equation.
The first limiting condition is that 
$|m_{AF}|$ is sufficiently larger than the kinetic energy 
${\cal O}(t\sqrt{x })$.
The second condition is 
$|v| \ll |m_{AF}|$ or $\sqrt{4t^2+m_{AF}^2} \ll |v|$.
With these conditions, the low-energy physics would be approximately described 
by quasi-particles of linear dispersions
at ${\vec k} = (\pi/2, \pi/2)$ and $(\pi/2, 0)$, respectively.
It is noted here that these quasi-particles are
not simple two-component Dirac fermions,
because the vertical stripe formation mixes
the two-component Dirac fermions with different chiralities
as mentioned above.
For the case $|v| < |v_c|\sim 2\pi xt^2/m_{AF}$,
the analysis is complicated because 
we should consider two bands doped in different way.
It is noted here that there are two upper bands and two lower bands
for each spin degree of freedom when $m_{AF}\neq 0$ and $v\neq 0$.
The analysis of $R_H$ in this case 
is given in the appendix~\ref{sec:special-case}.
When $|v|$ is larger than the critical value $|v_c|$,
only the second band is doped.
Therefore, for $|v_c|< |v| \ll |m_{AF}|$ or $\sqrt{4t^2+m_{AF}^2} \ll |v|$,
the result is expressed as
\begin{equation}
R_H \simeq
\frac{a^2}{|e|x}\cdot\frac{4t^2\tilde{t}^2}
{(t^2+\tilde{t}^2)^2},
\label{eq:RH-stripe}
\end{equation}
where 
$\tilde{t}=t\sqrt{1-|v/m_{AF}|}$ for $|v_c|< |v| \ll |m_{AF}|$,
or $\tilde{t}=t\sqrt{|v|/\sqrt{4t^2+m_{AF}^2}-1}$ 
for $\sqrt{4t^2+m_{AF}^2} \ll |v|$.

On the other hand, the thermopower 
$S_{\mu} \sim \beta\langle J^e_\mu  J^e_\mu \rangle^{-1}
\langle J^e_\mu  J^Q_\mu\rangle$, where $\vec{J}^e$ is 
the electric current density and $\vec{J}^Q$ is the heat current density, 
along the direction $\mu=x,y$ remains almost isotropic 
because the anisotropy of the correlation functions 
$\langle J^e_\mu  J^e_\mu\rangle$ and 
$\langle J^e_\mu  J^Q_\mu\rangle$ cancels each other in the numerator 
and denominator, respectively. However, $S$ is rather sensitive to 
the change of the gap and the electronic dispersion, 
because $\langle J^e_\mu  J^Q_\mu\rangle$ contains 
additional dimension of energy and is suppressed by 
the coexistence of the AF order and vertical (horizontal) stripe. 
For the case $|v_c|< |v| \ll |m_{AF}|$ or $\sqrt{4t^2+m_{AF}^2} \ll |v|$,
it is given at low temperature by
\begin{equation}
S \simeq
\frac{k_B}{|e|}\cdot \frac{\pi^2}{3}k_B T\cdot
\frac{8\pi x \,t\tilde{t}+2\tilde{m}^2}
{8\pi x \,t \tilde{t}
\left[8\pi x \,t \tilde{t}+ \tilde{m}^2
\right]^{\frac12}},
\label{eq:S-stripe}
\end{equation}
where $\tilde{t}=t\sqrt{1-|v/m_{AF}|}$ and $\tilde{m}=|m_{AF}|-|v|$
for $|v_c|< |v| \ll |m_{AF}|$,
or $\tilde{t}=t\sqrt{|v|/\sqrt{4t^2+m_{AF}^2}-1}$ 
and $\tilde{m}=|v|-\sqrt{4t^2+m_{AF}^2}$
for $\sqrt{4t^2+m_{AF}^2} \ll |v|$.
Comparing eq.~(\ref{eq:S-SFP}) in the case$\phi=\pi$ and eq.(22),
we can see that the weak stripe, i.e. $|v|\sim |v_c|$, 
strongly suppresses the thermopower 
with sufficiently strong AF order, i.e., large $|m_{AF}|$. 
Therefore a crucial test of this scenario is to measure 
the thermopower $S$ in the untwined sample where $R_H$ is not suppressed.  

\begin{figure*}
  \begin{tabular}{cc}
    \includegraphics[scale=0.5]{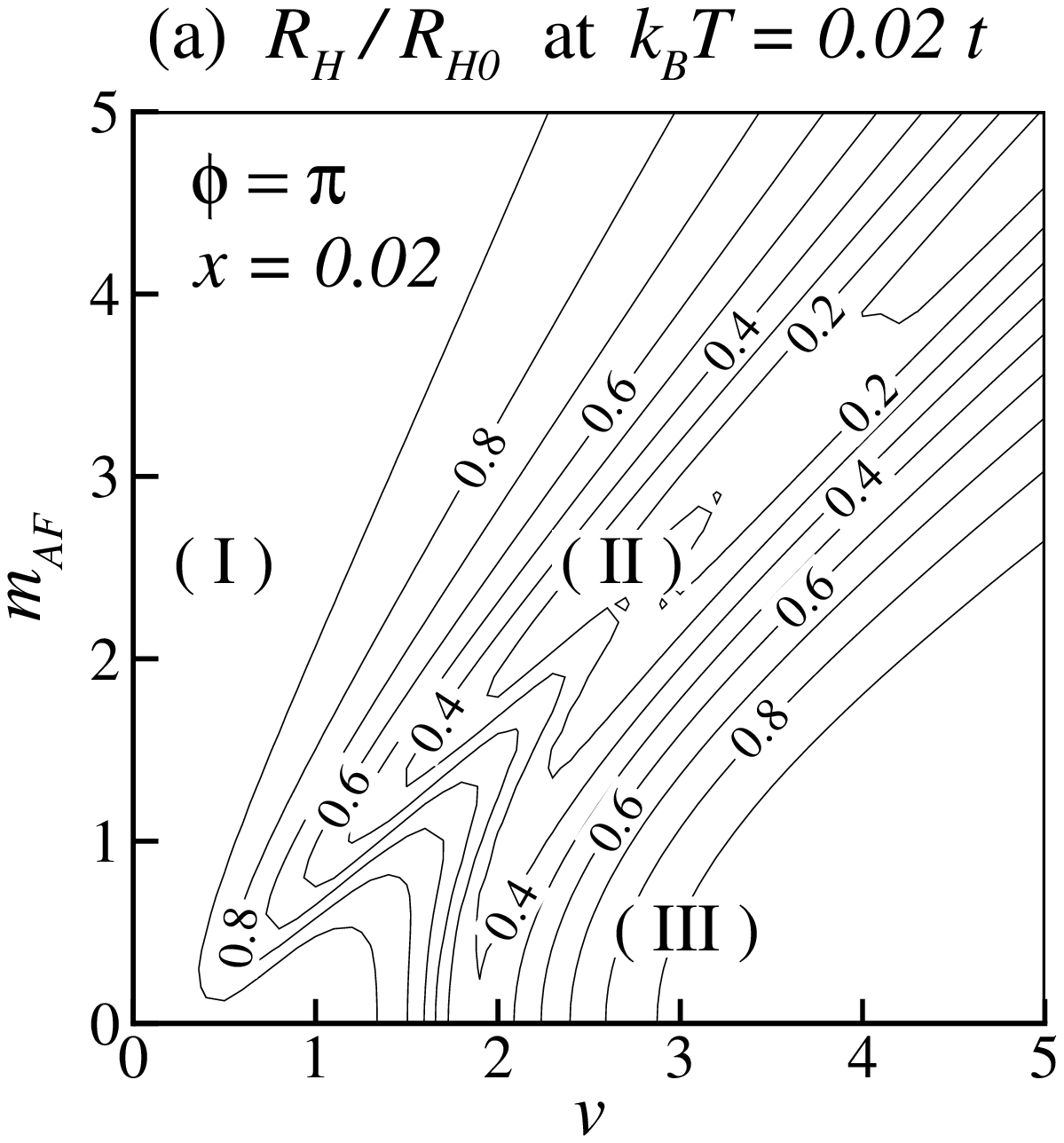} &
    \includegraphics[scale=0.5]{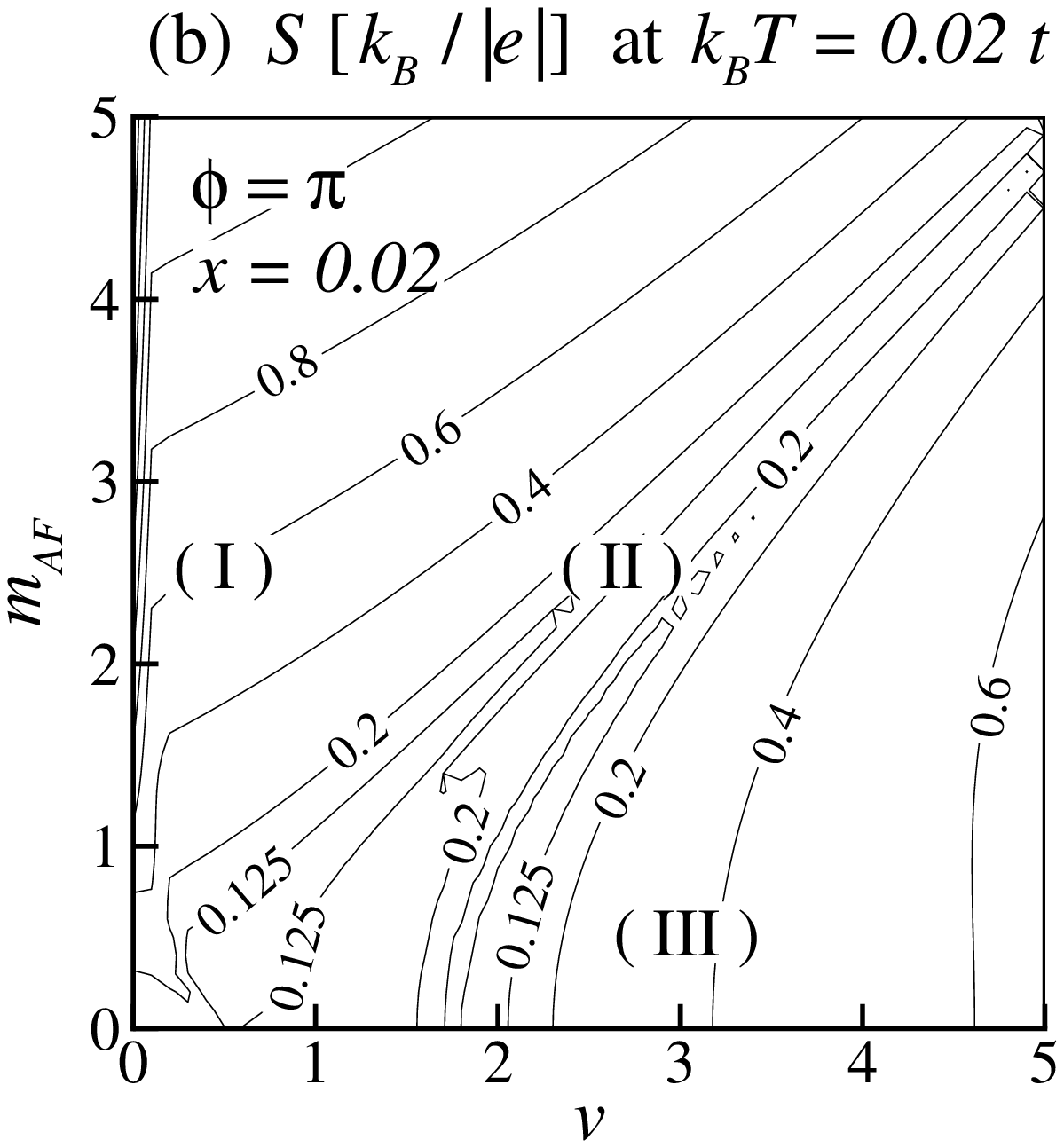} 
  \end{tabular}
  \caption{
(a) Hall coefficient $R_H$ divided by $R_{H0} = a^2/(|e|x)$, and 
(b) thermopower $S$ in the unit $k_B/|e|$.
Symbols (I), (II), and (III) correspond to
Figs.~\ref{fig:FS}(I), \ref{fig:FS}(II), and \ref{fig:FS}(III).
}
  \label{fig:RH-S}
\end{figure*}
\begin{figure*}
  \begin{tabular}{ccc}
    \includegraphics[scale=0.45]{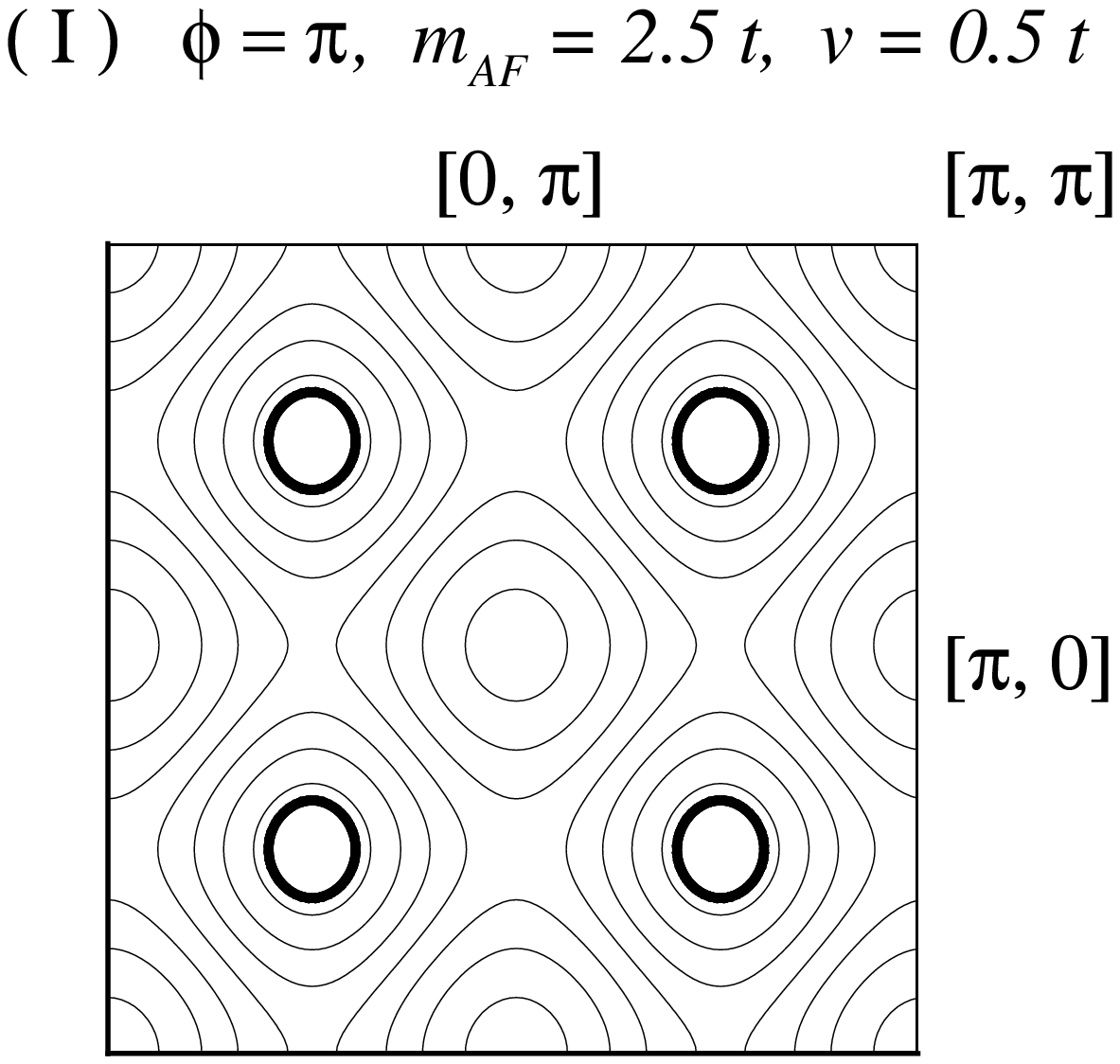} &
    \includegraphics[scale=0.45]{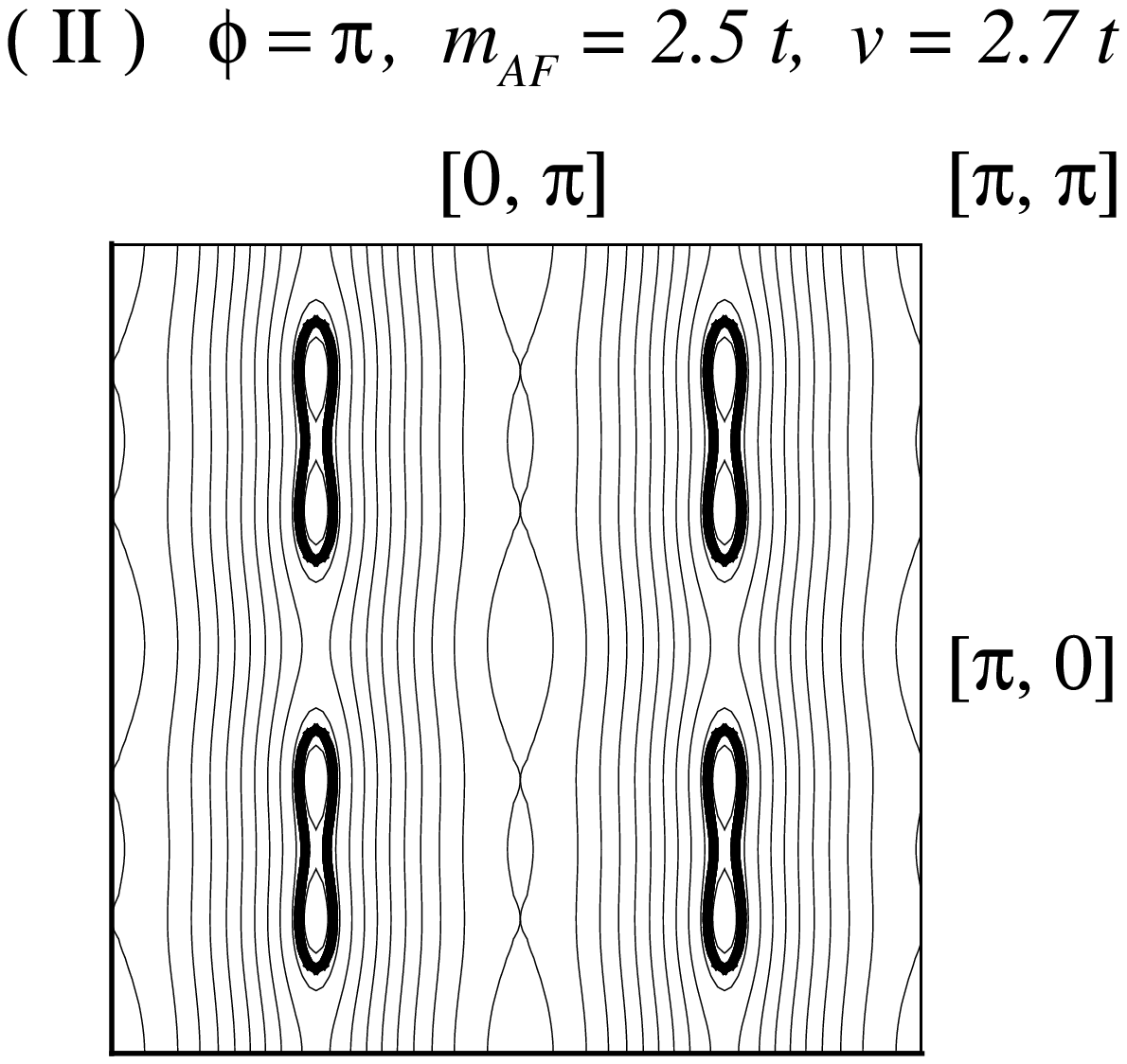} &
    \includegraphics[scale=0.45]{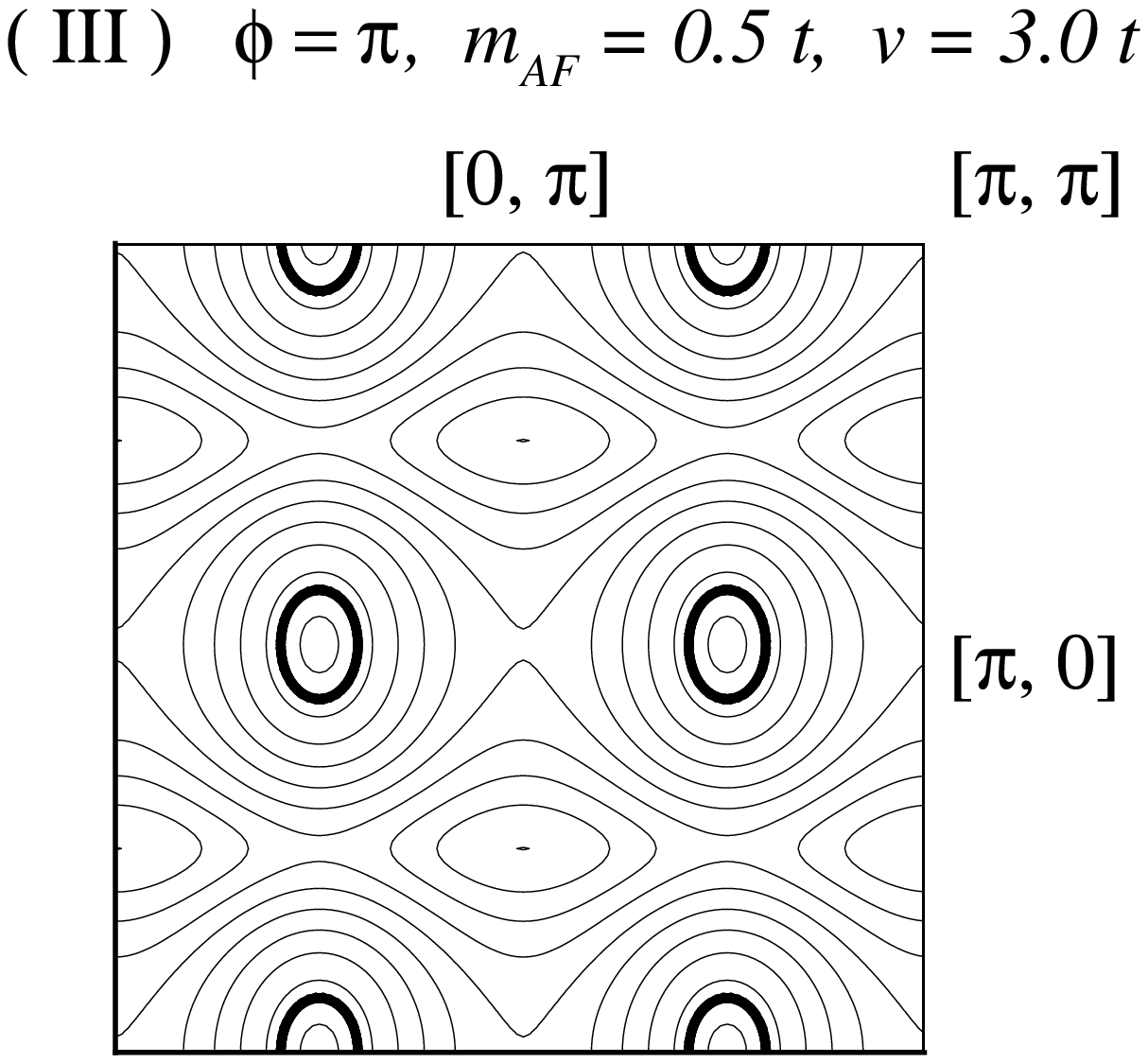} 
  \end{tabular}
  \caption{
Schematic view of the energy contours
for three parameter sets.
Thick lines in each figure represent
the Fermi surface for $x=0.02$.
The reduced Brillouin zone is the square 
with the corners $[\pm\pi/2, \pm\pi/2]$.
Copies are also drawn over the original Brillouin zone
for $\phi=m_{AF}=v=0$.
}
\label{fig:FS}
\end{figure*}
Now let us turn to the results for $R_H$ and $S$ in general cases. 
Although the analytic results for  $R_H$ 
(eq.~(\ref{eq:RH-stripe}))  and $S$  (eq.~(\ref{eq:S-stripe})) 
are valid only in the limiting cases,
they show what are crucial for each quantity.
As for $R_H$, the anisotropy of the transfer integral, 
i.e., the anisotropy of the velocity, is crucial with the help of
the averaging of the conductivity tensor.
On the other hand, as for $S$, the mass reduction, i.e., 
the reduction of the band gap, is also crucial 
as well as the anisotropy of the transfer integral.
Therefore, interesting results are expected in
the nontrivial case $|m_{AF}| \alt |v| \alt \sqrt{4t^2+m_{AF}^2}$,
because the energy dispersion is highly anisotropic
and there is no band gap between the two middle bands.
We employ the Boltzmann transport theory and numerically evaluated 
the correlation functions. Figs.~\ref{fig:RH-S}(a) and \ref{fig:RH-S}(b) 
shows $R_H$ and $S$ as a function of the strength of the AF order 
$m_{AF}$ and the vertical stripe $v$, respectively. 
The temperature is fixed at $0.02 t/k_B$. 
As mentioned above, the region near the left axis in
Fig.~\ref{fig:RH-S}(b) shows that
$S$ is once suppressed by the weak stripe, i.e., $|v|\sim |v_c|$.
Then both are remarkably suppressed 
when $m_{AF}$ and $v$ are comparable with each other, i.e., 
near the diagonal line. 
Figs.~\ref{fig:FS}(I), \ref{fig:FS}(II) and \ref{fig:FS}(III) 
show the equal-energy contours and Fermi surfaces 
for the 3 points in Figs.~\ref{fig:RH-S}(a) and \ref{fig:RH-S}(b).
Especially, Fig.~\ref{fig:FS}(II) belongs to
the nontrivial case and shows the peculiar shape of the Fermi surface.
According to ref.~\cite{geometric}, $\sigma_{xy}$ could be reduced 
in the nontrivial Fermi surface compared to that of the circular Fermi surface
surrounding the same size of area in $\vec{k}$-space. 
Especially when the Fermi surface have both parts of positive convexity 
and negative one, this reduction would be very effective 
as seen in the region $|v|\sim |m_{AF}|$.

Finally, the effects of the AF order and the vertical (horizontal) stripe
on the temperature dependences of $R_H$ and $S$ are shown 
in Figs.~\ref{fig:RH-S.tmp.u-v}(a) and \ref{fig:RH-S.tmp.u-v}(b).
The AF order enhances the both quantities at the almost 
all temperature range except for the region near the peak of $R_H$,
and shift the peaks to higher temperature.
This is because the AF order introduces the additional energy scale,
i.e., the band gap.
Then the vertical (horizontal) stripe remarkably suppresses 
both of $R_H$ and $S$ in the whole temperature range
when $v$ is comparable with $m_{AF}$.
These results are parallel to the above results
where the temperature is fixed.

\section{\label{sec:discussion}Discussion}
Now we discuss the above results in comparison with the experiments. 
The most important issue is whether the high temperature phase 
above $T_N$ and the pseudo-gap region can be described 
in terms of the massless Dirac fermions by putting $m_{AF} = 0$. 
The temperature and $x$-dependences of $R_H$ and 
$S$ appear to be qualitatively consistent with the experiments. 
However, when one looks at the $x$-dependence of 
the conductivity $\sigma_{xx}$ at $T = 300K$ above $T_N$, 
it fits better with $\propto x^{3/2}$  
rather than $\propto x$ due to the $x$-dependence 
of the hole mobility\cite{ando1}.
On the other hand,  $\sigma_{xx} \propto \sqrt{x}$ and 
$\sigma_{xy} \propto x^0$ in the our model, which is contradicting 
with the above experiments. Furthermore, the onset of
the AF order at $T_N$ does not affect 
the conductivity $\sigma_{xx}$ in the experiment\cite{ando1}, 
which is more difficult to understand from any mean field
picture. This is related to the interpretation of the insulating gap 
at $x=0$. In the mean field picture, it is due to the AF ordering while 
there remains a large gap even without the AF order in the Mott
insulator picture. If the large gap disappears and crossing of the two 
bands occurs above $T_N$, the upper band becomes relevant at around 
$T_x \cong J \sqrt{x}$ and particle-hole symmetry will be recovered 
above $T_x$.  This massless 
Dirac spectrum also gives the novel $x$-dependence as described above. 
Therefore it seems that the $x$-dependence observed in experiments 
suggests that the Mott insulating picture is more appropriate for the 
high temperature phase.  
In other words, the Dirac fermion without the AF gap is never
relevant to the underdoped cuprates.
This is also consistent with the asymmetry of ARPES 
between hole-doped\cite{ARPES} and electron-doped\cite{NCCO} cuprates. 
Recent experiments on NCCO\cite{NCCO} strongly suggest
that the minima of the electron dispersion are at 
${\vec k} = (\pi,0)$ and $(0,\pi)$, and the particle-hole symmetry is 
broken. This appears to be consistent with the SDW picture with
appropriate longer range hopping integrals $t'$, and $t''$.
However, the sign of $t'$ and $t''$ is reversed when one  
consider the $t$-$t'$-$t''$-$J$ model for electron doped case,  
and the minimum at $(\pi,0)$ is recovered by self-consistent Born 
approximation\cite{scba}, and its results can be interpreted as
the flux state with the AF order\cite{tklee}.
Therefore, it makes sense to consider the flux state together with
the AF order, but does not without it.

The AF ordered state, on the other hand, can be well described in 
terms of the mean field state with the flux order. 
In this AF state, the conventional behavior of the doped carriers are 
expected without the stripe formation. Therefore the effects of 
the stripes are the most interesting issue.
It is found that $R_H = a^2/(|e|x)$ is a rather robust feature
at low temperature. 
The suppression occurs only when the AF order and 
the vertical (horizontal) stripe coexist,
and the directional average is taken within the plane. 
The thermopower is also suppressed by the stripe formation, 
and does not need the directional average. 
The difference between ref.~\cite{PH-symmetry}
and ref.~\cite{ando1} seems to be  due to the sample preparation. 
The longer time annealing has been done for the former case,
while the sample is quenched in the latter case. Therefore it is
expected that the chain is more ordered in the former case, which
might help the stripe formation. The 4-fold symmetry observed in 
the magnetoresistance\cite{YBCO-stripe-exp} is interpreted
to be  due to the induced stripe in terms of the in-plane external 
magnetic field.

In conclusions, we have studied the transport properties 
of the flux state as a model for underdoped cuprates.
This model shows several remarkable features
such as parity anomaly, scaling laws for $R_H$ and $S$,
the recovery of particle-hole symmetry at high temperature.
Compared with the existing experimental data, 
in particularly the $x$-dependence of $\sigma_{xx}$,
it is unlikely that this model describes 
the underdoped cuprates without the AF order.
However, the flux state with the AF order,
which gives the mass gap to the Dirac fermions,
describes well the ordered state. In this case, the stripe
order affects the transport properties in a nontrivial way,
and we have discussed the experiments from this view point.

\section*{Acknowledgement}
The authors would like to acknowledge fruitful discussions with 
Y.~Ando, C.~M.~Ho, T.~K.~Lee, S.~Miyasaka, T.~Okuda, N.~P.~Ong, 
Y.~Tokura, S.~Uchida.
N.~N. is supported by Priority Areas Grants and Grant-in-Aid for COE research  
from the Ministry of Education, Science, Culture and  Sports of Japan. 

\appendix
\section{\label{sec:conductivity}Conductivity in the continuum model}
This appendix is devoted to present 
the uniform conductivity $\sigma_{\mu\nu}(\omega)$ of the continuum
effective theory constructed in sec.~\ref{sec:hamiltonian}.
When the effect of the relaxation is approximated by
replacing $\omega$ by $\Omega=\omega+i/\tau$
with the life time $\tau$ and
we can diagonalize a given Hamiltonian,
the linear response theory generally gives the following representation
for the uniform conductivity.
\begin{eqnarray}
\sigma_{\mu\nu}(\Omega)
&=& i\,e^2\sum_{\alpha,\beta}
\frac{f_F(\epsilon_\beta-\mu)-f_F(\epsilon_\alpha-\mu)}
{\epsilon_\alpha-\epsilon_\beta}
\nonumber\\
&&\qquad\times
\frac{\langle\alpha|J_{\nu}|\beta\rangle
\langle\beta|J_{\mu}|\alpha\rangle}
{\Omega+\epsilon_\beta-\epsilon_\alpha},
\end{eqnarray}
where $\alpha$ and $\beta$ are the quantum indices of eigen states,
$J_{\mu}$ represents $\mu$-component of the current density,
and $f_F(\epsilon-\mu)$ is the Fermi distribution function.
We apply the above formula to the effective theory 
in sec.~\ref{sec:hamiltonian},
i.e. (2+1)D Dirac fermions in the external magnetic field.
For example, in the case of 
the contribution from $1$-sector with $\sigma$-spin,
the following replacements are sufficient.
\begin{eqnarray}
&&\sum_{\alpha,\beta}\rightarrow 
\frac{|eB|}{2\pi}\sum_{n,n'=0}^{\infty}\sum_{s,s'=\pm},
\quad
|\alpha\rangle\rightarrow |n\,\sigma\,\pm\rangle,
\quad
\epsilon_{\alpha}\rightarrow \pm\epsilon_{n\sigma},
\nonumber\\
&&J_{x} \rightarrow \hat{j}_{x}=2ta\,\tau_{1},
\nonumber\\
&&J_{y} \rightarrow 
\hat{j}_{y}=2ta(\cos\frac{\phi}{2}\tau_{1}+\sin\frac{\phi}{2}\tau_{2}).
\end{eqnarray}
Here $|n\,\sigma\,\pm\rangle$ and $\epsilon_n$ are defined 
by eqs.~(\ref{eq:eigenvector})-(\ref{eq:zero-mode})
in sec.~\ref{sec:hamiltonian}.
$\tau_i$ are the Pauli matrices in
the space of $|\tilde{\uparrow}(\tilde{\downarrow})\rangle$.
[See eq.~(\ref{eq:eigenvector}).]
Then the contribution from $1$-sector with $\sigma$-spin is represented by
\begin{widetext}
\begin{eqnarray}
\sigma_{\mu\nu}^{1\sigma}(\Omega)
&=&
i\,e^2\frac{|eB|}{2\pi}
\sum_{n,n'=0}^{\infty}\sum_{s,s'=\pm}
\frac{f_F(s'\epsilon_{n'\sigma}-\mu)-f_F(s\epsilon_{n\sigma}-\mu)}
{s\epsilon_{n\sigma}-s'\epsilon_{n'\sigma}}
\frac{\langle n\,\sigma\,s|\hat{j}_{\nu}|n'\,\sigma\,s'\rangle
\langle n'\,\sigma\,s'|\hat{j}_{\mu}|n\,\sigma\,s\rangle}
{\Omega+s'\epsilon_{n'\sigma}-s\epsilon_{n\sigma}}
\nonumber\\
&=&
i\, e^2\frac{|eB|}{2\pi}
\sum_{n, n'=0}^{\infty}\sum_{s=\pm}\Biggl[
\frac1{\epsilon_{n'\sigma}+\epsilon_{n\sigma}}
\frac{\langle n\,\sigma\,s|\hat{j}_{\nu}|n'\,\sigma\,-s\rangle
\langle n'\,\sigma\,-s|\hat{j}_{\mu}|n\,\sigma\,s\rangle}
{\Omega-s(\epsilon_{n'\sigma}+\epsilon_{n\sigma})}
\nonumber\\
&&+
\frac{f_F(\epsilon_{n'\sigma}-s\mu)-f_F(\epsilon_{n\sigma}-s\mu)}
{\epsilon_{n\sigma}-\epsilon_{n'\sigma}}
\frac{\langle n\,\sigma\,s|\hat{j}_{\nu}|n'\,\sigma\,s\rangle
\langle n'\,\sigma\,s|\hat{j}_{\mu}|n\,\sigma\,s\rangle}
{\Omega+s(\epsilon_{n'\sigma}-\epsilon_{n\sigma})}
\nonumber\\
&&-
\frac{f_F(\epsilon_{n'\sigma}+s\mu)-f_F(\epsilon_{n\sigma}-s\mu)}
{\epsilon_{n\sigma}+\epsilon_{n'\sigma}}
\frac{\langle n\,\sigma\,s|\hat{j}_{\nu}|n'\,\sigma\,-s\rangle
\langle n'\,\sigma\,-s|\hat{j}_{\mu}|n\,\sigma\,s\rangle}
{\Omega-s(\epsilon_{n'\sigma}+\epsilon_{n\sigma})}\Biggr]
\nonumber\\
&=&
i\, e^2\frac{|eB|}{2\pi}
\sum_{n, n'=0}^{\infty}\sum_{s=\pm}\Biggl[
\frac{2f_F(\epsilon_{n\sigma}-s\mu)-1}
{\Omega^2-(\epsilon_{n'\sigma}+\epsilon_{n\sigma})^2}
\nonumber\\
&&\times
\Biggl\{
\frac{(-i\Omega)\Re\left[\langle n\,\sigma\,s|\hat{j}_{\nu}|n'\,\sigma\,-s\rangle
\langle n'\,\sigma\,-s|\hat{j}_{\mu}|n\,\sigma\,s\rangle\right]}
{\epsilon_{n'\sigma}+\epsilon_{n\sigma}}
+s\,\Im\left[\langle n\,\sigma\,s|\hat{j}_{\nu}|n'\,\sigma\,-s\rangle
\langle n'\,\sigma\,-s|\hat{j}_{\mu}|n\,\sigma\,s\rangle\right]
\Biggr\}
\nonumber\\
&&+
\frac{2f_F(\epsilon_{n\sigma}-s\mu)}
{\Omega^2-(\epsilon_{n'\sigma}-\epsilon_{n\sigma})^2}
\nonumber\\
&&\times
\Biggl\{
\frac{(-i\Omega)
\Re\left[\langle n\,\sigma\,s|\hat{j}_{\nu}|n'\,\sigma\,s\rangle
\langle n'\,\sigma\,s|\hat{j}_{\mu}|n\,\sigma\,s\rangle\right]}
{\epsilon_{n\sigma}-\epsilon_{n'\sigma}}
+s\,\Im\left[\langle n\,\sigma\,s|\hat{j}_{\nu}|n'\,\sigma\,s\rangle
\langle n'\,\sigma\,s|\hat{j}_{\mu}|n\,\sigma\,s\rangle\right]
\Biggr\}\Biggr]
\end{eqnarray}
\end{widetext}
It is noted that all distribution functions in the first line represent 
those of electrons.
In the second line, we have done the transformation, 
$f_F(-\epsilon-\mu)\rightarrow 1-f_F(\epsilon+\mu)$, 
for negative energy modes
in order to change the distribution function of the valence band
to that of the quasi-hole.
In the third line, the difference of distribution functions
is pulled apart to each distribution function.
This procedure is justified because
$\langle n\,\sigma\,s|\hat{j}_{\mu}|n'\,\sigma\,s'\rangle\neq0$ 
only for $n' = n \pm 1$ as we will see below
and the distribution function $f_F(\epsilon_{n\sigma}\pm\mu)$
decreases sufficiently fast as $n$ increases.

The final procedure is to substitute the matrix elements
by the following explicit forms,
\begin{widetext}
\begin{eqnarray}
\langle n\,\sigma\,\pm|\tau_{1}|n'\,\sigma\,\pm\rangle
&=&\pm\frac12\left[
\sqrt{\frac{E_{n\sigma\pm}E_{n+1\sigma\mp}}
{\epsilon_{n\sigma}\epsilon_{n+1\sigma}}}
\delta_{n+1,n'}
+
\sqrt{\frac{E_{n'+1\sigma\mp}E_{n'\sigma\pm}}
{\epsilon_{n'+1\sigma}\epsilon_{n'\sigma}}}
\delta_{n,n'+1}
\right],
\nonumber\\
\langle n\,\sigma\,\pm|\tau_{1}|n'\,\sigma\,\mp\rangle
&=&\mp\frac12\left[
\sqrt{\frac{E_{n\sigma\pm}E_{n+1\sigma\pm}}
{\epsilon_{n\sigma}\epsilon_{n+1\sigma}}}
\delta_{n+1,n'}
-
\sqrt{\frac{E_{n'+1\sigma\mp}E_{n'\sigma\mp}}
{\epsilon_{n'+1\sigma}\epsilon_{n'\sigma}}}
\delta_{n,n'+1}
\right],
\nonumber\\
\langle n\,\sigma\,\pm|\tau_{2}|n'\,\sigma\,\pm\rangle
&=&\mp i\,\mbox{sgn}(eB_{\phi})\frac12\left[
\sqrt{\frac{E_{n\sigma\pm}E_{n+1\sigma\mp}}
{\epsilon_{n\sigma}\epsilon_{n+1\sigma}}}
\delta_{n+1,n'}
-
\sqrt{\frac{E_{n'+1\sigma\mp}E_{n'\sigma\pm}}
{\epsilon_{n'+1\sigma}\epsilon_{n'\sigma}}}
\delta_{n,n'+1}
\right],
\nonumber\\
\langle n\,\sigma\,\pm|\tau_{2}|n'\,\sigma\,\mp\rangle
&=&\pm i\,\mbox{sgn}(eB_{\phi})\frac12\left[
\sqrt{\frac{E_{n\sigma\pm}E_{n+1\sigma\pm}}
{\epsilon_{n\sigma}\epsilon_{n+1\sigma}}}
\delta_{n+1,n'}
+
\sqrt{\frac{E_{n'+1\sigma\mp}E_{n'\sigma\mp}}
{\epsilon_{n'+1\sigma}\epsilon_{n'\sigma}}}
\delta_{n,n'+1}
\right]
\end{eqnarray}
\end{widetext}
where $E_{n\sigma\pm}=\epsilon_{n\sigma}\pm\mbox{sgn}(eB_{\phi})u_{\sigma}$.
(The absence of one of zero modes is appropriately represented
by $E_{n\sigma\pm}$.)
As for the dc conductivity,
the contribution from $1$-sector with $\sigma$-spin is given by
\begin{widetext}
\begin{eqnarray}
\sigma_{xx}^{1\sigma}
&=&
\frac{e^2}{2\pi}\,\tau^{-1}\,K_{eB}^2
\sum_{s=\pm}\Biggl[
f_F(|u_{\sigma}|-s\mu)\frac{|u_{\sigma}|[1+\mbox{sgn}(su_{\sigma}eB_\phi)]}
{(K_{eB_\phi}^2+\tau^{-2})^2+4\tau^{-2}u_{\sigma}^2}
\nonumber\\
&&+
\sum_{n=1}f_F(\epsilon_n-s\mu)
\left[(K_{eB_\phi}^4-\tau^{-4})^2
+8(K_{eB_\phi}^4+\tau^{-4})\tau^{-2}\epsilon_n^2
+(4\tau^{-2}\epsilon_n^2)^2\right]^{-1}
\nonumber\\
&&\quad \times
\left[\tau^{-2}(\tau^{-2}+4\epsilon_n^2)
(\epsilon_n+u_{\sigma}^2\epsilon_n^{-1})
+K_{eB_\phi}^4(3\epsilon_n-u_{\sigma}^2\epsilon_n^{-1})
-4K_{eB_\phi}^2\tau^{-2}|u_{\sigma}|\mbox{sgn}(su_{\sigma}eB_\phi)\right]
\Biggr]
\nonumber\\
& &+\frac{e^2}{2\pi}\,\frac{\tau^{-1}\,K_{eB}^2}{2}\sum_{n=0}^{\infty}
\frac{1}{(\epsilon_n+\epsilon_{n+1})[\tau^{-2}+(\epsilon_n+\epsilon_{n+1})^2]}
\left(1+\frac{u_{\sigma}^2}{\epsilon_n\epsilon_{n+1}}\right),
\label{eq:sigma_xx}
\\
\sigma_{xy}^{1\sigma}
&=&
\frac{e^2}{2\pi}\,\mbox{sgn}(eB)\,K_{eB_\phi}^2
\sum_{s=\pm}\Biggl[
\frac{s}{2}\,f_F(|u_{\sigma}|-s\mu)\frac{(K_{eB_\phi}^2+\tau^{-2})
[1+\mbox{sgn}(su_{\sigma}eB_\phi)]}
{(K_{eB_\phi}^2+\tau^{-2})^2+4\tau^{-2}u_{\sigma}^2}
\nonumber\\
&&+
\sum_{n=1}s\,f_F(\epsilon_n-s\mu)
\left[(K_{eB_\phi}^4-\tau^{-4})^2
+8(K_{eB_\phi}^4+\tau^{-4})\tau^{-2}\epsilon_n^2
+(4\tau^{-2}\epsilon_n^2)^2\right]^{-1}
\nonumber\\
&&\quad \times
\left[K_{eB_\phi}^2(K_{eB_\phi}^4-\tau^{-4}+4\tau^{-2}u_{\sigma}^2)
+\tau^{-2}(K_{eB_\phi}^4-\tau^{-4}+4\tau^{-2}\epsilon_n^2)
|u_{\sigma}|\epsilon_n^{-1}\mbox{sgn}(su_{\sigma}eB_\phi)\right]
\Biggr]
\nonumber\\
& &-\frac{e^2}{2\pi}\,
\mbox{sgn}\left(\sin\frac{\phi}{2}\right)\frac{K_{eB_\phi}^2}{2}
\sum_{n=0}^{\infty}
\frac{u_{\sigma}(\epsilon_n+\epsilon_{n+1})}
{\epsilon_n\epsilon_{n+1}[\tau^{-2}+(\epsilon_n+\epsilon_{n+1})^2]}
\nonumber\\
&&+\sigma_{xx}^{1\sigma}\cos\frac{\phi}{2},
\label{eq:sigma_xy}
\end{eqnarray}
\end{widetext}
The terms without the distribution function represent 
the parts of the inter-band effect
which remain even in the case $\mu=0$ and $k_{B}T=0$.
As for $\sigma_{xx}^{1\sigma}$, the contribution from the inter-band effect 
is negligible in the semi-classical limit 
($K_{eB_{\phi}}\ll 1 \ll |\mu|\tau$)
compared to the contribution from the Fermi level.
In this case, the Boltzmann theory is good approximation.
(The inter-band effect is neglected in the Boltzmann theory.)
On the other hand, as for $\sigma_{xy}^{1\sigma}$,
the last two terms, which include the inter-band effect
remaining in the case $\mu=0$ and $k_{B}T=0$,
cancel out after summing up the contribution from all sectors.
In other words, when the parity symmetry is breaking,
i.e. the numbers of right and left Dirac fermions are unbalanced,
we can not neglect the remaining inter-band effect.
Especially in the case $\tau=\infty$ and $B\to 0$,
the last line but one in $\sigma_{xy}^{1\sigma}$ gives 
a contribution $\pm 1/2$ in the unit $e^2/h$ as long as $u_{\sigma}\neq 0$,
where the sign depends on $u_{\sigma}$ and $\phi$,
and $h$ is Planck's constant.
However, it is noted that the total $\sigma_{xy}$
should take a integer value in the unit $e^2/h$ when it is quantized.
Therefore, in the case where this inter-band effect is crucial, 
we must seriously consider the contribution from 
the bottom of valence bands and the top of conduction bonds, 
which are out of range of the continuum model.

\section{\label{sec:mu-tmp}Temperature dependence of the chemical potential}
Here we consider the temperature dependence of the chemical potential $\mu$
in the limit $B\to 0$.
When there is no stripe formation,
the doping parameter of the continuum model is given by
\begin{eqnarray}
-x 
&=& d_{s} \sum_{c=\pm}\sum_{s=\pm}\int \frac{d^2 p}{(2\pi)^2}
\;s\,f_{F}(\epsilon_{c}(\vec{p})-s\mu)
\nonumber\\
&=&\frac{d_{f}}{8\pi t^2\left|\sin\frac{\phi}{2}\right|} 
\sum_{s=\pm}\int^\infty_{|m_{AF}|}
d\epsilon\;\epsilon\,s\,f_{F}(\epsilon-s\mu).
\label{eq:doping}
\end{eqnarray}
Here $d_{s}$ is the number of spin degrees of freedom, i.e. $d_{s} =2$,
$d_{f}$ is the total number of inner degrees of freedom, i.e. 
$d_{f} = (\mbox{left}+\mbox{right})\times d_{s} =4$, and
\begin{equation}
\epsilon_{\pm}(\vec{p})=\sqrt{(2ta)^2
\left(p^2\pm2\cos\frac{\phi}{2}\,p_{x}p_{y}\right)+m_{AF}^2}.
\end{equation}
The sign of $x$ is taken as it is positive when $\mu < 0$.

In the low-temperature approximation, $k_{B}T\ll(|\mu|-|m_{AF}|$,
we can estimate the right hand side of eq.~(\ref{eq:doping}) by using
the sharpness of the Fermi distribution function as follows,
\begin{eqnarray}
-x&=&\frac{d_{f}}{16\pi t^2}\,\mbox{sgn}(\mu)\, \theta(|\mu|-|m_{AF}|)
\nonumber\\
&&\quad\times
\left[\mu^2-m_{AF}^2+\frac{\pi^2}{3}(k_{B}T)^2+\cdots\right].
\end{eqnarray}
Then, in the hole doping case, i.e. $\mu < -|m_{AF}|$,
the temperature dependence of $\mu$ is given by
\begin{equation}
|\mu| \cong 
\sqrt{4\pi x t^2 \left|\sin\frac{\phi}{2}\right|+m_{AF}^2
-\frac{\pi^2}{3}(k_{B}T)^2},
\label{eq:mu-low}
\end{equation}
where we have used $d_{f}=4$.

On the other hand, in the high-temperature approximation, $k_{B}T\gg|\mu|$,
we can estimate  the right hand side of eq.~(\ref{eq:doping}) as follows
\begin{eqnarray}
-x 
&=&\frac{d_{f}}{8\pi t^2\left|\sin\frac{\phi}{2}\right|} 
\sum_{s=\pm} \frac1{\beta^2}\int^\infty_{\beta |m_{AF}|}
dy\;\frac{s\,y}{1+e^{y-s\beta\mu}},
\nonumber\\
&=&\frac{d_{f}}{4\pi t^2\left|\sin\frac{\phi}{2}\right|}
\frac{\mu}{\beta}
\Biggl[\frac{\beta|m_{AF}|}{1+e^{\beta |m_{AF}|}}
\nonumber\\
&&\quad
+\ln(1+e^{-\beta |m_{AF}|})+ \cdots\Biggr],
\end{eqnarray}
where $\beta=1/(k_{B}T)$ and we have expanded
the integrand in $\beta\mu$.
Finally, in the hole doping case, i.e. $\mu < 0$, 
the temperature dependence of $\mu$ is given by
\begin{equation}
|\mu|\cong \frac{\pi x t^2\beta\left|\sin\frac{\phi}{2}\right|}
{\frac{\beta|m_{AF}|}{1+e^{\beta |m_{AF}|}}
+\ln(1+e^{-\beta|m_{AF}|})},
\label{eq:mu-high}
\end{equation}
where $d_{f}=4$ is substituted.
It is noted that, fixing the doping parameter and the temperature,
the above formula is a increasing function of $|m_{AF}|$.
Therefore, the AF order $m_{AF}$ suppresses the recovery of 
the particle-hole symmetry.

\section{\label{sec:scaling}Scaling laws of $R_H$ and $S$}
In this appendix, we present the derivation of
the scaling laws of $R_H$ and $S$
in the case of $m_{AF}=m_{\rm diag}=v=0$ and in the limit $B\to 0$.
In order to see the temperature dependence of
the Fermi distribution function $f_F(\xi)$,
we introduce the function $f(\beta \xi)=f_F(\xi)$ where $\beta=1/(k_{B}T)$.
Then, from eq.~(\ref{eq:doping}) with $m_{AF}=0$,
the doping parameter $x$ is given by the following equation,
\begin{eqnarray} 
-x&=&\frac1{2\pi t^2 \left|\sin\frac{\phi}{2}\right|}
\sum_{s=\pm}\int_{0}^{\infty} d\epsilon\;\epsilon\,s\,f(\beta(\epsilon-s\mu))
\nonumber\\
&=&\frac1{2\pi t^2\left|\sin\frac{\phi}{2}\right|}
\sum_{s=\pm}\frac1{\beta^2}\int_{0}^{\infty} dy\;y\,s\,f(y-s\beta\mu),
\nonumber\\
\end{eqnarray}
It is easy to see that $t\sqrt{x}/(k_{B}T)$
is represented by a function of $\mu/(k_{B}T)$.
Therefore, when we can consider the inverse of the function,
$\mu/(k_{B}T)$ is represented by a function of $t\sqrt{x}/(k_{B}T)$.

In the same way, by making the energy integral dimensionless,
the response functions are represented as
\begin{widetext}
\begin{eqnarray}
\sigma_{xx}
&\cong&
\frac{e^2}{2\pi\left|\sin\frac{\phi}{2}\right|}
\left(\frac{2\tau}{\beta}\right)
\sum_{s=\pm}
\int_{0}^{\infty}dy
\Biggl[
-y\frac{df}{dy}(y-s\beta\mu)
+\left(\frac12-f(y-s\beta\mu)\right)
\frac1{1+\left(\frac{2\tau}{\beta}y\right)^2}
\Biggr],
\\
\sigma_{xy}
&\cong&
\frac{e^2}{2\pi}\left|\sin\frac{\phi}{2}\right|
\frac{2 a^2 eB}{x}(\beta t\sqrt{x})^2
\left(\frac{2\tau}{\beta}\right)^2
\sum_{s=\pm} 
\int_{0}^{\infty}dy
\,s
\Biggl[-\frac{df}{dy}(y-s\beta\mu)\Biggr]
\Biggl[1+
\frac1{1+\left(\frac{2\tau}{\beta}y\right)^2}
\Biggr],
\\
\beta\,\sigma_{xx}^Q
&\cong&
\frac{e}{2\pi\left|\sin\frac{\phi}{2}\right|}
\left(\frac{2\tau}{\beta}\right)
\sum_{s=\pm}
\int_{0}^{\infty}dy
\Biggl[
-s\, y (y-s\beta\mu)\frac{df}{dy}(y-s\beta\mu)
-\left(\frac12-f(y-s\beta\mu)\right)
\frac{\beta\mu}{1+\left(\frac{2\tau}{\beta}y\right)^2}
\Biggr],
\end{eqnarray}
\end{widetext}
where $\sigma_{\mu\nu}^Q$ is the response function given by 
the correlation function of the electric current density $e\vec{J}$
and the heat current density $\vec{J}^Q$.
In the first-quantized representation,
the heat current density
for the $1$-sector with $\sigma$-spin is given by
\begin{eqnarray}
\hat{j}^Q_{x}&=&
(2ta)^2\left(\hat{p}_{x}+\cos\frac{\phi}{2}\,\hat{p}_{y} \right)
- \mu \hat{j}_{x},
\\
\hat{j}^Q_{y}&=&
(2ta)^2\left(\hat{p}_{y}+\cos\frac{\phi}{2}\,\hat{p}_{x} \right)
- \mu \hat{j}_{y}.
\end{eqnarray}
This is the continuum version of the heat current density
derived in the lattice model,
which is presented in the appendix~\ref{sec:heat-current}.
The right hand side of each response function is
a function of $t\sqrt{x}/(k_{B} T)$ and $\tau k_{B}T$,
because $\mu/(k_{B} T)$ is a function of $t\sqrt{x}/(k_{B} T)$.
Therefore, $x\,R_{H}\cong x\,\sigma_{xy}/(B \sigma_{xx}^2)$
and $S\cong\sigma^Q_{xx}/(T\sigma_{xx})$ is
functions of $t\sqrt{x}/(k_{B} T)$ and $\tau k_{B}T$.
In the semi-classical limit, the dependence of $\tau k_{B}T$
is negligible, and $x\,R_H$ and $S$ scale
as functions of $t\sqrt{x}/(k_{B} T)$.

\section{\label{sec:heat-current}Heat current density in the lattice model}
The definition of the heat current density
is not straightforward as against that of the electric current density.
Here we present the heat current density
of the system represented by eq.~(\ref{eq:Hamiltonian}).
The Hamiltonian is rewritten as $H=\sum_{\vec{r}}h_{\vec{r}}$ where
\begin{eqnarray}
h_{\vec{r}}
&=&\frac12\sum_{\hat{\delta},\sigma}
\left[
-t_{\vec{r}+\hat{\delta},\vec{r}}\;
c^{\dagger}_{\vec{r}+\hat{\delta},\sigma}c_{\vec{r}\sigma}
-t^{*}_{\vec{r}+\hat{\delta},\vec{r}}\;
c^{\dagger}_{\vec{r}\sigma}c_{\vec{r}+\hat{\delta},\sigma}
\right]
\nonumber\\
&&+\sum_{\sigma}(V_{\vec{r}\sigma}-\mu)
c^{\dagger}_{\vec{r}\sigma}c_{\vec{r}\sigma},
\\
V_{\vec{r}\sigma}&=&
-\left[u_{\sigma}\cos(\vec{Q}\cdot\vec{r})
+v\cos(\vec{Q}_v\cdot\vec{r})\right],
\end{eqnarray}
and the sum of the unit vector $\hat{\delta}$ runs $\pm\hat{x}$
and $\pm\hat{y}$.
It is clear that $h_{\vec{r}}$ is
interpreted as a local heat density.
The heat current density would be conceptually
defined by the averaging the product
of the local velocity and $h_{\vec{r}}$.
However, it is difficult to define the local velocity,
especially in the second quantized formalism.
Here we define the heat current density
from the analogy with the conceptual definition.
First, using $h_{\vec{r}}$, we introduce the quantity $\vec{R}$ as follows,
$
\vec{R}=V^{-1}\sum_{\vec{r}}\vec{r}\;h_{\vec{r}},
$
where $V$ is the volume (area) of the system.
This quantity has the dimension of 
heat density multiplied by length.
Then, we define the heat current density 
as a time derivative of $\vec{R}$, i.e.
$\vec{J}^{Q}=i[H,\vec{R}]$.
From the dimension of $\vec{R}$,
$\vec{J}^{Q}$ correctly has the dimension of heat current density.
Finally, the explicit form of $\vec{J}^Q$ is given by
\begin{widetext}
\begin{eqnarray}
\vec{J}^{Q}
&=& \vec{J}^{E}-\mu \vec{J},
\\
\vec{J}^{E}
&=&\frac{1}{V}\sum_{\vec{r}}
\left[
-\frac{i}{2}\sum_{\hat{\delta},\hat{\delta}',\sigma}
(\hat{\delta}+\hat{\delta}')\;
t_{\vec{r}+\hat{\delta},\vec{r}}\;
t^{*}_{\vec{r}+\hat{\delta},\vec{r}}\;
c^{\dagger}_{\vec{r}+\hat{\delta}+\hat{\delta}',\sigma}c_{\vec{r}\sigma}
+\frac{i}{2}\sum_{\hat{\delta},\sigma}
\hat{\delta}\;
(V_{\vec{r}+\hat{\delta}}+V_{\vec{r}})\;
t_{\vec{r}+\hat{\delta},\vec{r}}\;
c^{\dagger}_{\vec{r}+\hat{\delta},\sigma}c_{\vec{r}\sigma}
\right],
\nonumber\\
&=&\frac1{V}\sum_{\vec{k},\sigma}
\Biggl[
\sum_{\hat{\mu}=\hat{x},\hat{y}}
(-4t^2)\;\hat{\mu}\;\sin k_{\mu}
\left\{
\cos k_{\mu}+\cos\frac{\phi}{2}\cos(\epsilon_{\mu\nu}k_\nu)
\right\}
c^{\dagger}_{\vec{k}\sigma}c_{\vec{k}\sigma}
\nonumber\\
&&\qquad\qquad
+2vt\;\hat{y}\;\sin k_{y}
\left\{
-\cos\frac{\phi}{4}c^{\dagger}_{\vec{k}\sigma}c_{\vec{k}+\vec{Q}_{x},\sigma}
+i\sin\frac{\phi}{4}c^{\dagger}_{\vec{k}\sigma}c_{\vec{k}+\vec{Q}_{y},\sigma}
\right\}
\Biggr],
\\
\vec{J}
&=&
\frac{i}{V}\sum_{\vec{r}}
\sum_{\hat{\delta},\sigma}\hat{\delta}\;
t_{\vec{r}+\hat{\delta},\vec{r}}\;
c^{\dagger}_{\vec{r}+\hat{\delta},\sigma}c_{\vec{r}\sigma}
=
\frac1{V}\sum_{\vec{k},\sigma}
\sum_{\hat{\mu}=\hat{x},\hat{y}}
2t\;\hat{\mu}\;\sin k_{\mu}
\left[
\cos\frac{\phi}{4}
c^{\dagger}_{\vec{k}\sigma}c_{\vec{k}\sigma}
+(-1)^{\mu}i \sin\frac{\phi}{4}
c^{\dagger}_{\vec{k}\sigma}c_{\vec{k}+\vec{Q},\sigma}
\right],
\end{eqnarray}
\end{widetext}
where $\vec{Q}=(\pi,\pi)$, $\vec{Q}_{x}=(\pi,0)$, and $\vec{Q}_{y}=(0,\pi)$.
The symbol $(-1)^{\mu}$ means that $(-1)^x=1$ and $(-1)^y=-1$.
In a physical sense, $\vec{J}^{E}$ is the energy current density, 
and $\vec{J}$ is the current density.

As for the continuum limit,
the following transformation of bases
are performed before taking the limit.
\begin{equation}
\left[
\begin{array}{l}
c_{\vec{k}\sigma}\\
c_{\vec{k}+\vec{Q},\sigma}\\
c_{\vec{k}+\vec{Q}_x,\sigma}\\
c_{\vec{k}+\vec{Q}_y,\sigma}
\end{array}
\right]
=
\left[
\begin{array}{cc}
U_{+} & O \\
O & U_{-}
\end{array}
\right]
\left[
\begin{array}{c}
\mbox{\boldmath $\psi$}_{\vec{k},1\sigma}\\
\mbox{\boldmath $\psi$}_{\vec{k},2\sigma}
\end{array}
\right],
\end{equation}
where $\mbox{\boldmath $\psi$}_{\vec{k},i\sigma}$ ($i=1,2$)
is a two-component operator respectively, and 
\begin{equation}
U_{\pm} 
= \frac1{\sqrt{2}}(\tau_{3}\pm\tau_{1})
\left(
\cos\frac{\phi}{8}\,\tau_{1}\pm\sin\frac{\phi}{8}\,\tau_{2}
\right).
\end{equation}
When there is no stripe,
$\mbox{\boldmath $\psi$}_{\vec{k},i\sigma}$
with $\vec{k}\sim(\pi/2,\pi/2)$
corresponds to the Dirac fermion of $i$-sector ($i=1,2$)
with $\sigma$-spin.

\section{\label{sec:special-case} $R_H$ in the case $|v| < |v_c|$}
Here we present the analysis of $R_H$ 
when two lower bands are doped.
Around the tops of these two bands,
their dispersions are approximated as
\begin{equation}
\epsilon_{\pm}(\vec{p})
=-\sqrt{(2ta)^2 p_x^2+(2\tilde{t}_{\pm}a)^2 p_y^2+\tilde{m}_{\pm}},
\end{equation}
where
$\tilde{t}_{\pm}=t\sqrt{1\pm|v/m_{AF}|}$ and
$\tilde{m}_{\pm}=|m_{AF}|\pm|v|$.
In this case, by considering the two bands separately
and then adding the contributions at the stage of conductivity,
we can get the Hall coefficient $R_H\cong a^2/(|e|x)$ at low temperature,
where $x$ is the total doping parameter.
However, when there are domains of stripes with different directions, 
we should take an average of conductivity tensor.
Then, we obtain the following result.
\begin{equation}
R_H \cong
\frac{a^2}{|e|}\cdot
\frac{4\left[x-\left|\frac{v}{m_{AF}}\right|(x_{-}-x_{+})\right]}
{\left[2x-\left|\frac{v}{m_{AF}}\right|(x_{-}-x_{+})\right]^2},
\end{equation}
where $x_{+}$ and $x_{-}$ are doping parameters
for the first and second bands respectively.
Then, from the conservation of the total doping $x$ and
the commonness of the chemical potential $\mu$,
$x_{+}$ and $x_{-}$ are determined by the following equations,
\begin{eqnarray}
&&x_{-}+x_{+} = x,
\\
&&x_{-}\sqrt{1-\left|\frac{v}{m_{AF}}\right|}
-x_{+}\sqrt{1+\left|\frac{v}{m_{AF}}\right|}
= \frac{|v\, m_{AF}|}{2\pi t^2}.
\nonumber\\
\end{eqnarray}
It is noted here that $x_{\pm}$ includes 
the contribution of both spin degrees of freedom. 
When $|v|$ is larger than the critical value $|v_c|$, i.e., 
\begin{widetext}
\begin{equation}
|v| > |v_c|=|m_{AF}|\left(\frac{\sqrt{2\pi x }\,t}{m_{AF}}\right)^2
\left[
\sqrt{1+\frac14\left(\frac{\sqrt{2\pi x }\,t}{m_{AF}}\right)^4}
-\frac12\left(\frac{\sqrt{2\pi x }\,t}{m_{AF}}\right)^2
\right],
\end{equation}
\end{widetext}
only the second band is doped, i.e., $x_{+}=0$ and $x_{-}=x$.

We can also analyze the thermopower $S$ by the similar way.
However, its expression is more complicated than that of $R_H$ 
and is not suggestive.
Therefore, we do not give the explicit analysis,
but pointed out that, in the region $0 < |v| <|v_c|$, 
$S$ is highly reduced when $|v|$ increases
as shown in Fig. \ref{fig:RH-S}(b).

\end{document}